\documentclass[12pt,preprint]{aastex}
\usepackage{graphicx}

\begin{document}
\title{Reconstructing the Primordial Spectrum from WMAP Data \\
       by the Cosmic Inversion Method}

\author{Noriyuki Kogo\altaffilmark{1}, Makoto Matsumiya\altaffilmark{2}, 
        Misao Sasaki\altaffilmark{3}, and Jun'ichi Yokoyama\altaffilmark{4}}

\affil{\altaffilmark{1,2,4}Department of Earth and Space Science, 
       Graduate School of Science, Osaka University, \\
       Toyonaka 560-0043, Japan}
\affil{\altaffilmark{3}Yukawa Institute for Theoretical Physics, 
       Kyoto University, Kyoto 606-8502, Japan}

\email{\altaffilmark{1}kogo@vega.ess.sci.osaka-u.ac.jp, 
       \altaffilmark{2}matumiya@vega.ess.sci.osaka-u.ac.jp, \\
       \altaffilmark{3}misao@yukawa.kyoto-u.ac.jp, 
       \altaffilmark{4}yokoyama@vega.ess.sci.osaka-u.ac.jp}

\begin{abstract}
We reconstruct the primordial spectrum of the curvature perturbation, $P(k)$, 
from the observational data of the Wilkinson Microwave Anisotropy Probe (WMAP) 
by the cosmic inversion method developed recently. 
In contrast to conventional parameter-fitting methods, 
our method can potentially reproduce small features in $P(k)$ 
with good accuracy. 
As a result, we obtain a complicated oscillatory $P(k)$. 
We confirm that this reconstructed $P(k)$ recovers 
the WMAP angular power spectrum with resolution up to $\Delta \ell \simeq 5$. 
Similar oscillatory features are found, however, 
in simulations using artificial cosmic microwave background data generated 
from a scale-invariant $P(k)$ with random errors that mimic observation. 
In order to examine the statistical significance of the nontrivial features, 
including the oscillatory behaviors, therefore, 
we consider a method to quantify the deviation from scale-invariance 
and apply it to the $P(k)$ reconstructed from the WMAP data. 
We find that there are possible deviations from scale-invariance 
around $k \simeq 1.5\times10^{-2}$ and $2.6\times10^{-2}{\rm Mpc}^{-1}$. 
\end{abstract}

\keywords{cosmic microwave background---cosmology: theory}

\section{Introduction} \label{INTRO}
The Wilkinson Microwave Anisotropy Probe (WMAP) satellite has brought us 
interesting information about our universe \citep{WMAPBASIC}. 
From the remarkably precise observation of 
the temperature fluctuations and the polarization of 
the cosmic microwave background (CMB), 
one can obtain not only accurate values of the global cosmological parameters, 
but also invaluable information of the properties on 
the primordial fluctuations \citep{WMAPBASIC,WMAPPARA,WMAPINF,WMAPGAUSS}. 
Although their results as a whole support 
the standard $\Lambda$CDM (cold dark matter) model 
with Gaussian, adiabatic, and scale-invariant primordial fluctuations, 
some features that cannot be explained by the standard model 
have also been pointed out, 
such as (1) lack of power on large scales, 
(2) running of the spectral index $n_s$ 
from $n_s>1$ on larger scales to $n_s<1$ on smaller scales, 
and (3) oscillatory behaviors of the power spectrum on intermediate scales. 

In fact, feature 1 was known already with COBE data \citep{COBE}, 
and a possible explanation was proposed \citep{JY99}. 
There have been many new proposals 
\citep{BFZ03,DCS03,GE03A,URLW04,CPKL03,CCL03,FZ03,KT03,DK03,PFZ03}, 
as well as many arguments about its statistical significance 
\citep{WMAPPARA,TDH03,DTZH03,GWMMH03,NJP04,GE03B,GE04}. 
A number of inflation models that can account for feature 2 
have also been proposed recently 
\citep{FLZZ03,KS03,KYY03,HL03,CST03,YY03,DK03}. 

On the other hand, feature 3, namely, possible oscillatory behaviors 
around a simple power-law spectrum, is more difficult to quantify 
\citep{WMAPINF}. 
Although several attempts to reconstruct 
the primordial spectrum were made by combining WMAP data with 
other independent observational data, 
such as the Two-Degree Field Galaxy Redshift Survey (2dFGRS) 
and Ly$\alpha$ forest data 
\citep{WSS99a,WSS99b,SH01,SH03,TZ02,WM02,BLWE03,MW03A,MW03B,MW03C}, 
they all employed the binning, wavelet band powers, 
or direct wavelet expansion method to the data. 
These methods cannot detect possible oscillatory behaviors 
if their scale is smaller than the binning scale. 
It is preferable to use a method that can restore the primordial spectrum 
as a continuous function without any ad hoc filtering scale 
to investigate detailed features such as 3. 
Fortunately, such a new method 
has been proposed in \citet{MSY02} and \citet{MSY03}, 
with the name cosmic inversion method, 
and their test calculations using mock data without observational errors 
have shown that this method can reproduce possible small dips and peaks off 
a scale-invariant spectrum quite well, 
in the spatially flat universe with the adiabatic initial condition. 
In this paper, we attempt to reconstruct $P(k)$ from 
the WMAP data using this method. 
To apply it to the actual data, 
we consider the observational errors by Monte Carlo simulations 
and use a modified CMBFAST\footnote{See http://www.cmbfast.org/} code 
that adopts much finer resolution than the original one 
in both $k$ and $\ell$, so that 
we can compute the fine structure of angular power spectra accurately. 
We mention that recently another new method for the reconstruction of $P(k)$ 
has been proposed by \citet{SS03}. 

This paper is organized as follows. 
In Sec.~\ref{METHOD}, we review the cosmic inversion method 
and describe our analysis method using the WMAP data. 
In Sec.~\ref{RESULTS}, we apply our method to the WMAP data 
and discuss the results. 
In particular, we investigate statistical significance of 
deviations from scale-invariance.
Finally, we present our conclusion in Sec.~\ref{CONCLUSION}. 

\section{Inversion Method} \label{METHOD}
First, we briefly review the cosmic inversion method 
proposed by \citet{MSY02,MSY03}. 
We assume a spatially flat universe with an adiabatic initial condition, 
both of which are generic predictions of standard inflation 
and have been supported by the WMAP data. 
In the end, we obtain a first-order differential equation 
for the primordial spectrum $P(k)$. 

The CMB anisotropy is quantified by the angular correlation function 
defined as 
\begin{eqnarray}
C(\theta)
\equiv \left\langle \Theta(\hat{{\bf n}}_1)
       \Theta(\hat{{\bf n}}_2) \right\rangle
     = \sum_{\ell=0}^{\infty} \frac{2\ell+1}{4\pi}C_\ell P_\ell(\cos\theta),
\quad \cos\theta=\hat{{\bf n}}_1 \cdot \hat{{\bf n}}_2 \, ,
\label{CR}
\end{eqnarray}
where $\Theta(\hat{{\bf n}})$ is the temperature fluctuation in 
the direction $\hat{{\bf n}}$. 
We decompose the Fourier components of the temperature fluctuations 
$\Theta(\eta,k)$ into multipole moments, 
\begin{eqnarray}
\Theta(\eta,k,\mu)
=\sum_{\ell=0}^{\infty} (-i)^\ell \Theta_\ell(\eta,k) P_\ell(\mu),
\label{MULTI}
\end{eqnarray}
where $\mu \equiv \hat{{\bf k}} \cdot \hat{{\bf n}}$, 
$k$ is the comoving wavenumber, 
and $\eta$ is the conformal time, with its present value being $\eta_0$. 
Using $\Theta_\ell(\eta,k)$, 
the angular power spectrum is expressed as 
\begin{eqnarray}
\frac{2\ell+1}{4\pi} C_\ell
=\frac{1}{2\pi^2} \int_0^\infty \! \frac{dk}{k} \,
 \frac{k^3 \bigl\langle |\Theta_\ell(\eta_0,k)|^2 \bigl\rangle}{2\ell+1}.
\label{CL}
\end{eqnarray}
The Boltzmann equation for $\Theta(\eta,k)$ 
can be transformed into the following integral form \citep{HS95}: 
\begin{eqnarray}
(\Theta+\Psi)(\eta_0,k,\mu)
=\int_0^{\eta_0} \! d\eta \, \Bigl\{
 [\Theta_0+\Psi-i\mu V_{{\rm b}}] {\cal V}(\eta)
 +(\dot{\Psi}-\dot{\Phi}) e^{-\tau(\eta)}
 \Bigl\} e^{ik\mu(\eta-\eta_0)},
\label{THETA}
\end{eqnarray}
where the overdot denotes the derivative with respect to the conformal time. 
Here, $\Psi$ and $\Phi$ are the Newton potential 
and the spatial curvature perturbation in the Newton gauge, 
respectively \citep{KS84}, and 
\begin{eqnarray}
{\cal V}(\eta) \equiv \dot{\tau} e^{-\tau(\eta)}, \quad
    \tau(\eta) \equiv \int_{\eta}^{\eta_0} \! \dot{\tau} d\eta,
\label{VIS&TAU}
\end{eqnarray}
are the visibility function and the optical depth for Thomson scattering, 
respectively. 
In the limit that 
the thickness of the last scattering surface (LSS) is negligible, we have 
${\cal V}(\eta) \approx \delta(\eta-\eta_*),\,
e^{-\tau(\eta)} \approx \theta(\eta-\eta_*)$, 
where $\eta_*$ is the recombination time 
when the visibility function is maximum \citep{HS95}. 
Taking the thickness of the LSS into account, 
we have a better approximation for Eq.~(\ref{THETA}) as 
\begin{eqnarray}
(\Theta+\Psi)(\eta_0,k,\mu)
\approx \int_{\eta_{*{\rm start}}}^{\eta_{*{\rm end}}} \! d\eta \, \Bigl\{
        [\Theta_0+\Psi-i\mu\Theta_1] {\cal V}(\eta)
        +(\dot{\Psi}-\dot{\Phi}) e^{-\tau(\eta)}
        \Bigl\} e^{-ik\mu d}
\equiv  \Theta^{{\rm app}}+\Psi,
\label{APPROX}
\end{eqnarray}
where $d \equiv \eta_0-\eta_*$ is the conformal distance 
from the present to the LSS 
and $\eta_{*{\rm start}}$ and $\eta_{*{\rm end}}$ are the times 
when the recombination starts and ends, respectively. 
Here we introduce the transfer functions, $f(k)$ and $g(k)$, defined by 
\begin{eqnarray}
\int_{\eta_{*{\rm start}}}^{\eta_{*{\rm end}}} \! d\eta \,
\left[ (\Theta_0+\Psi)(\eta,k){\cal V}(\eta)
      +(\dot{\Psi}-\dot{\Phi})(\eta,k) e^{-\tau(\eta)} \right]
&\equiv& f(k) \Phi(0,{\bf k}), \\
\int_{\eta_{*{\rm start}}}^{\eta_{*{\rm end}}} \! d\eta \,
\Theta_1(\eta,k){\cal V}(\eta)
&\equiv& g(k) \Phi(0,{\bf k})\,.
\label{TRANS}
\end{eqnarray}
We can calculate $f(k)$ and $g(k)$ numerically; 
they depend only on the cosmological parameters, 
for we are assuming that only adiabatic fluctuations are present. 
Then, we find the approximated multipole moments as 
\begin{eqnarray}
\Theta_\ell^{{\rm app}}(\eta_0,k)
=(2\ell+1) \left[ f(k) j_\ell(kd)+g(k) j'_\ell(kd) \right] \Phi(0,{\bf k}),
\label{TAPP}
\end{eqnarray}
and the approximated angular correlation function as 
\begin{eqnarray}
C^{{\rm app}}(r)
=\sum_{\ell=\ell_{{\rm min}}}^{\ell_{{\rm max}}} \frac{2\ell+1}{4\pi}
 C^{{\rm app}}_\ell P_\ell \left( 1-\frac{r^2}{2d^2} \right),
\label{CRAPP}
\end{eqnarray}
where $C^{{\rm app}}_\ell$ is obtained 
by putting Eq.~(\ref{TAPP}) into Eq.~(\ref{CL}), 
$r$ is defined as $r=2d \sin(\theta/2)$ on the LSS, 
and $\ell_{{\rm min}}$ and $\ell_{{\rm max}}$ 
are lower and upper bounds on $\ell$ 
due to the limitation of the observation. 
In the small-scale limit $r \ll d$, 
using the Fourier sine formula, 
we obtain a first-order differential equation 
for the primordial power spectrum of the curvature perturbation, 
$P(k) \equiv \langle |\Phi(0,{\bf k})|^2 \rangle$, 
\begin{eqnarray}
&&-k^2f^2(k)P'(k)+\left[ -2k^2f(k)f'(k)+kg^2(k) \right] P(k)
\nonumber\\
&&\qquad\qquad
=4\pi \int_0^\infty \! dr \,
 \frac{1}{r} \frac{\partial}{\partial r} \{ r^3 C^{{\rm app}}(r) \} \sin kr
 \equiv S(k).
\label{DIFF}
\end{eqnarray}
Since $f(k)$ and $g(k)$ are oscillatory functions around zero, 
we can find values of $P(k)$ at the zero points of $f(k)$ as 
\begin{eqnarray}
P(k_s)=\frac{S(k_s)}{k_s \, g^2(k_s)} \quad {\rm for} \quad f(k_s)=0,
\label{BC}
\end{eqnarray}
assuming that $P'(k)$ is finite at the singularities, $k=k_s$. 
If the cosmological parameters and the angular power spectrum are given, 
we can solve Eq.~(\ref{DIFF}) 
as a boundary value problem between the singularities. 

However, because Eq.~(\ref{DIFF}) is derived by adopting 
the approximation in Eq.~(\ref{APPROX}), 
$C^{{\rm app}}_\ell$ is different from 
the exact angular spectrum $C^{{\rm ex}}_\ell$ for the same initial spectrum. 
The errors caused by the approximation, 
or the relative differences between 
$C^{{\rm app}}_\ell$ and $C^{{\rm ex}}_\ell$, are as large as about 30\%. 
Thus, we should not use the observed power spectrum 
$C^{{\rm obs}}_\ell$ directly in Eq.~(\ref{CRAPP}). 
Instead, we must find the $C^{{\rm app}}_\ell$ that 
would be obtained for the actual $P(k)$. 
Of course, this is impossible in the rigorous sense, 
since it is the actual $P(k)$ that we are to reconstruct. 
It turns out to be possible, however, 
with accuracy high enough for our present purpose, 
because we find that the ratio, 
\begin{eqnarray}
b_\ell \equiv
\frac{C^{{\rm ex}}_\ell}{C^{{\rm app}}_\ell},
\label{RATIO}
\end{eqnarray}
is almost independent of $P(k)$ \citep{MSY03}. 
Using this fact, we first calculate the ratio, 
$b^{(0)}_\ell=C^{{\rm ex}(0)}_\ell/C^{{\rm app}(0)}_\ell$, 
for a known fiducial initial spectrum $P^{(0)}(k)$ 
such as the scale-invariant one. 
Then, inserting $C^{{\rm obs}}_\ell/b^{(0)}_\ell$, 
which is much closer to the actual $C^{{\rm app}}_\ell$, 
into the source term of Eq.~(\ref{DIFF}), 
we may solve for $P(k)$ with good accuracy. 

In practice, we cannot take the upper bound of the integration 
in the right-hand side of Eq.~(\ref{DIFF}) to be infinite. 
The integrand in Eq.~(\ref{DIFF}) is oscillating with 
its amplitude increasing with $r$. 
We therefore introduce a cutoff scale $r_{{\rm cut}}$. 
However, this inevitably introduces a smoothing scale to our method. 
As the cutoff scale is made larger, 
the rapid oscillations of the integrand with increasing amplitude 
become numerically uncontrollable. 
On the other hand, if the cutoff scale is made smaller, 
the resolution in the $k$-space becomes worse as 
$\Delta k \simeq \pi/r_{{\rm cut}}$. 
In the actual calculations, 
to maintain as good a numerical accuracy as possible 
and to obtain simultaneously as fine a resolution in $k$-space as possible, 
we convolve an exponentially decreasing function into the integration 
of the Fourier sine transform 
with the optimized cutoff scale of $r_{{\rm cut}} \simeq 0.5d$, 
corresponding to $\theta \simeq 30\degr$, or $\Delta\ell \sim 6$. 
Thus, the resolution of the Fourier sine transform is limited to 
$\Delta kd \simeq 6$. 
We could not adopt a finer resolution because of the numerical instability. 
Nevertheless, the resolution of our method is much finer than 
that of any other method. 
Note that there is an absolute theoretical limit 
$r_{\rm cut} \lesssim 2d$, or $\Delta kd \gtrsim 1.5$, because of 
the finite size of the LSS sphere. 

In this paper, we also account for the effect of 
observational errors on the reconstructed $P(k)$. 
First, for each $\ell$ we draw a random number from a Gaussian distribution 
whose mean value is equal to a central value of the observed $C_\ell$ 
and variance is given by the diagonal term of the covariance matrix, 
and then we reconstruct $P(k)$ from each simulated data set. 
In fact, each $C_\ell$ is weakly correlated with other multipoles 
and follows a $\chi^2$ distribution. 
However, our procedure is valid 
because the correlation with other multipoles is weak enough 
and the $\chi^2$ distribution is practically identical to 
the Gaussian distribution for sufficiently large $\ell$ where we analyze. 
We estimate the mean value and the variance of 
the reconstructed $P(k)$ at each $k$ for 1000 realizations. 
We find that the intrinsic errors caused by our inversion method itself, 
whose magnitude is estimated 
by inverting $C_\ell$ spectra calculated from artificial $P(k)$ spectra 
without observational errors, 
are much smaller than the observational errors of WMAP, 
except around the singularities where the numerical errors are amplified. 

To calculate $C^{{\rm ex}}_\ell$, 
we use a modified CMBFAST code with much finer resolutions 
than the original one in both $k$ and $\ell$. 
We limit $C_\ell$ in the range $20 \le \ell \le 700$ 
in order not to use data that have large observational errors 
due to the cosmic variance at small $\ell$ 
and the detector noise at large $\ell$. 
We adopt the fiducial initial spectrum, $P^{(0)}(k)$, 
as the scale-invariant spectrum, 
and the fiducial cosmological parameter set as 
$h=0.72$, $\Omega_b=0.047$, $\Omega_\Lambda=0.71$, 
$\Omega_m=0.29$, and $\tau=0.17$, 
which are the best-fit values to the WMAP data 
for the scale-invariant spectrum, $k^3P(k)=A$. 
In this case, the positions of the singularities given by Eq.~(\ref{BC}) 
are $kd \simeq 70, 430, 680, ...$, where 
$d \simeq 1.34\times10^4{\rm Mpc}$. 
Because the reconstructed $P(k)$ around the singularities has 
large numerical errors that are amplified by the observational errors, 
we can obtain $P(k)$ with good accuracy 
in the limited range $120 \lesssim kd \lesssim 380$ or 
$9.0\times10^{-3}{\rm Mpc}^{-1} \lesssim k 
\lesssim 2.8\times10^{-2}{\rm Mpc}^{-1}$. 
Then, we vary some cosmological parameters 
to examine how the shape of the reconstructed $P(k)$ is affected. 
That is, we also calculate for some cases of 
$h=0.65, 0.70, 0.75, 0.80$, $\Omega_b=0.03, 0.04, 0.05, 0.06$, 
and $\Omega_\Lambda=0.65, 0.70, 0.75, 0.80$, respectively. 
Note that since $\tau$ affects the shape of the power spectrum 
only on large scales, except for the normalization, 
we use a fixed value as $\tau=0.17$. 

\section{Results and Discussion} \label{RESULTS}

\subsection{Reconstruction from Original Data} \label{ORIGINAL}
\begin{figure}
\begin{center}
\includegraphics[width=10cm]{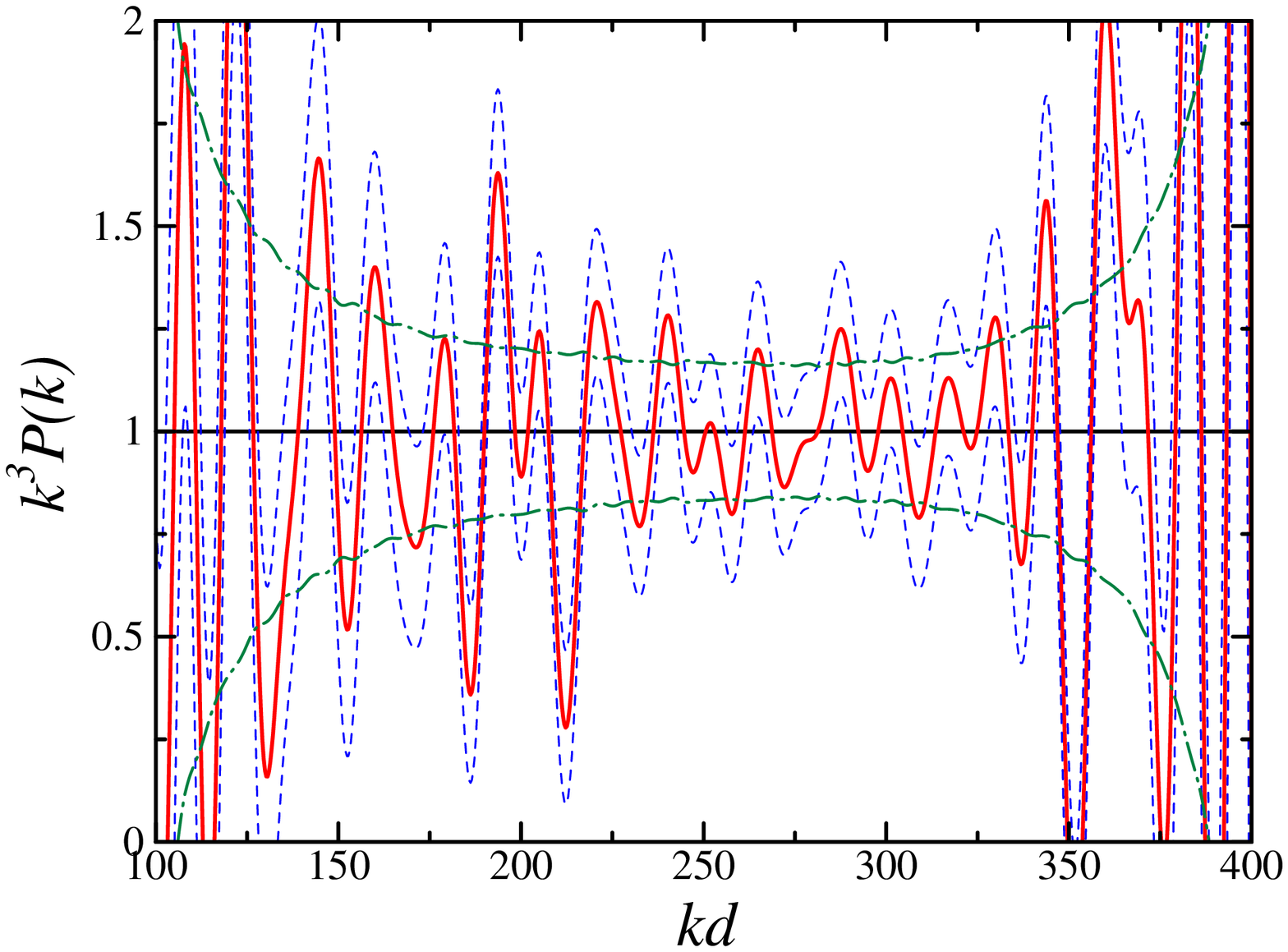}
\caption{Primordial spectrum $P(k)$ reconstructed from the WMAP data 
for $h=0.72$, $\Omega_b=0.047$, $\Omega_\Lambda=0.71$, 
$\Omega_m=0.29$, and $\tau=0.17$. 
The solid curve and the dashed curves represent 
the mean and the $1 \sigma$ errors, 
respectively, of the reconstructed $P(k)$, 
while the dash-dotted curves represent 
the $1 \sigma$ from the scale-invariance. 
The horizontal axis $kd$ corresponds roughly to $\ell$. 
The singularities lie at $kd \simeq 70$ and $430$. 
Some prominent features are seen around 
$kd \simeq 200$ and $350$. \label{REWMAP}}
\end{center}
\end{figure}

\begin{figure}
\begin{center}
\includegraphics[width=9cm]{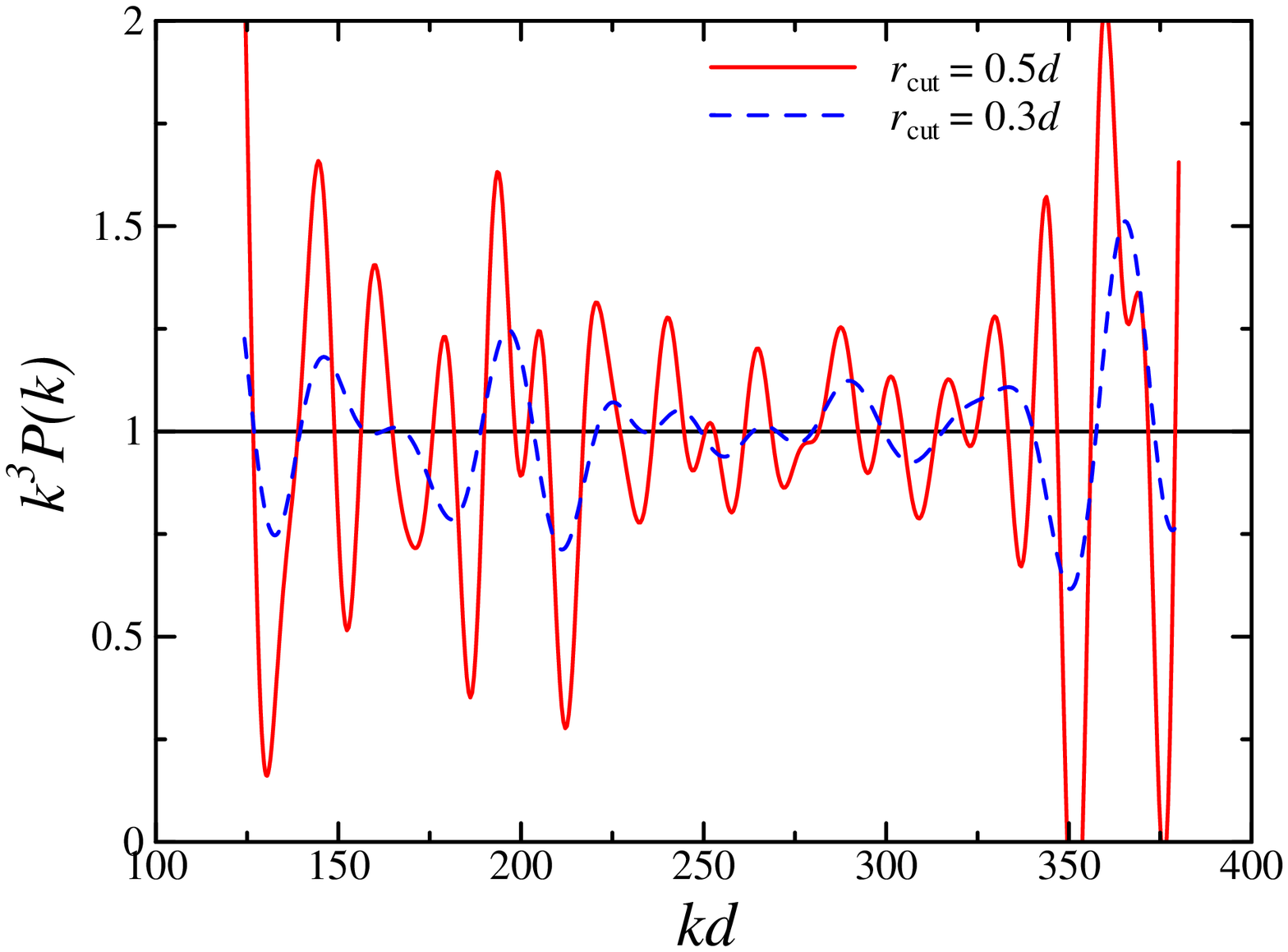}
\includegraphics[width=9cm]{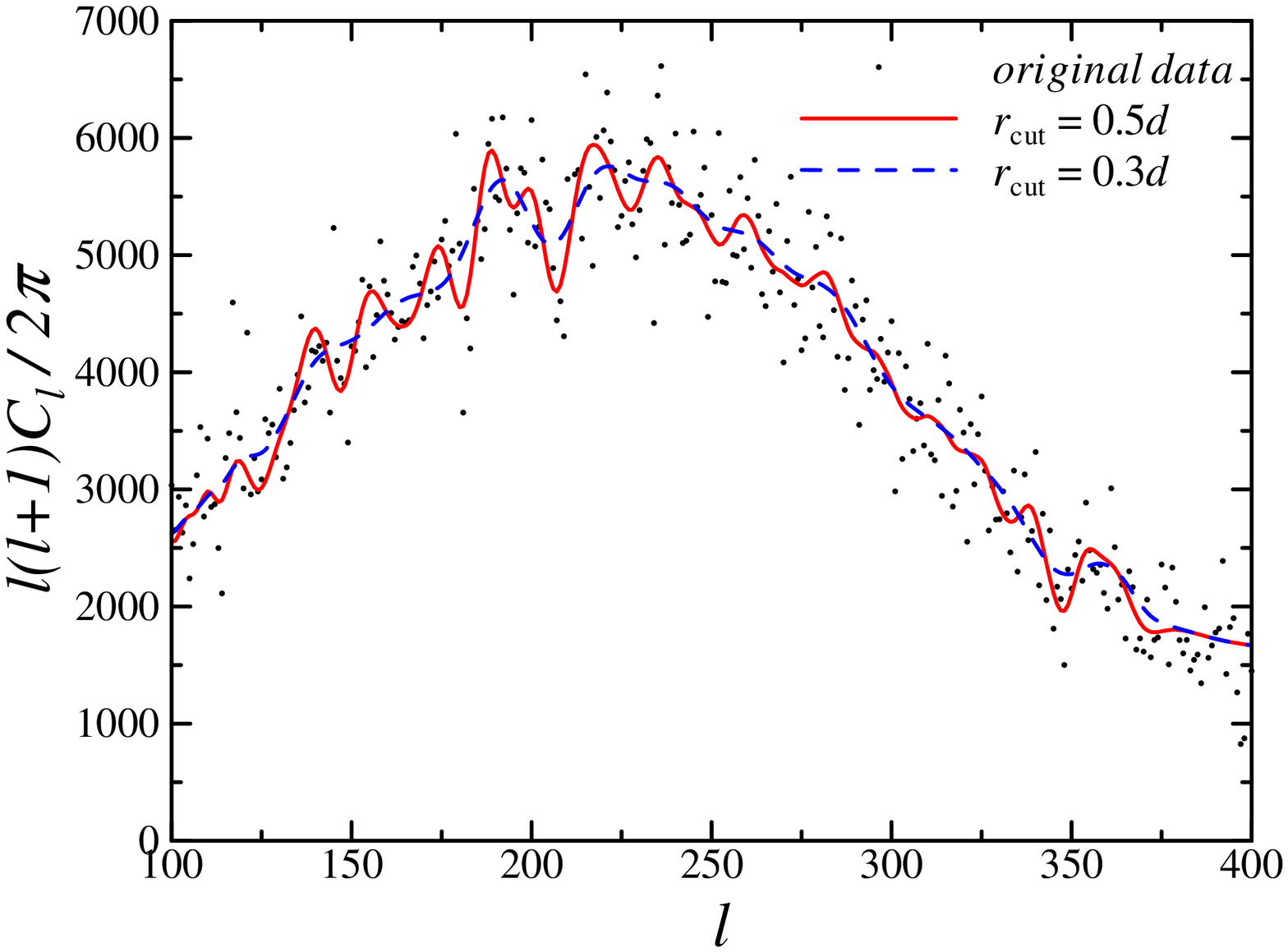}
\caption{Dependence on the cutoff scale $r_{{\rm cut}}$. 
{\it Top}: Reconstructed $P(k)$ spectra from the WMAP data for 
$r_{{\rm cut}} \simeq 0.5d$ and $0.3d$. 
We see that the resolution becomes better as the cutoff scale is made larger. 
{\it Bottom}: Recovered $C_\ell$ spectra 
from the obtained $P(k)$ spectra. \label{CUT}}
\end{center}
\end{figure}

\begin{figure}
\begin{center}
\includegraphics[width=8cm]{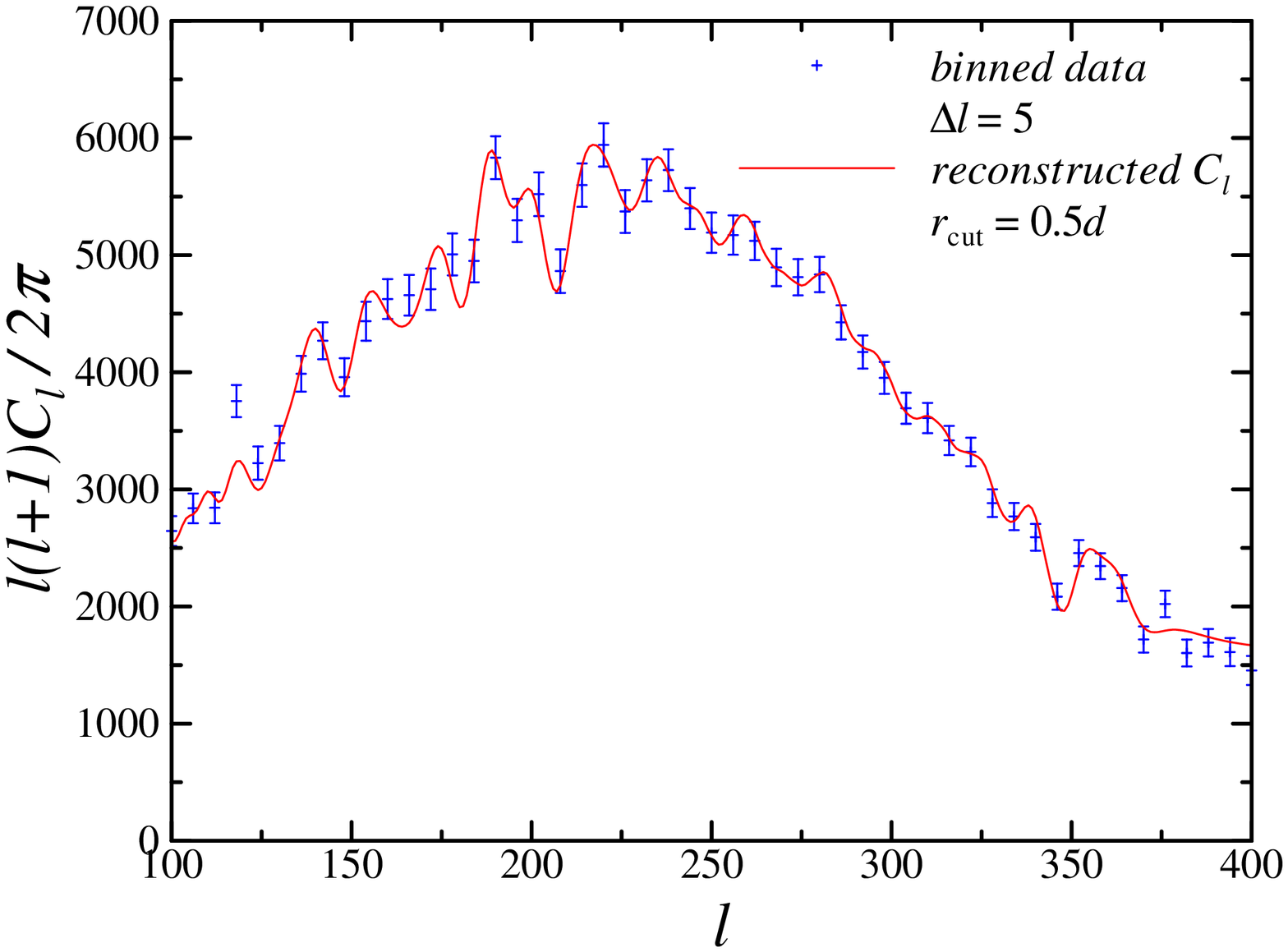}
\includegraphics[width=7cm]{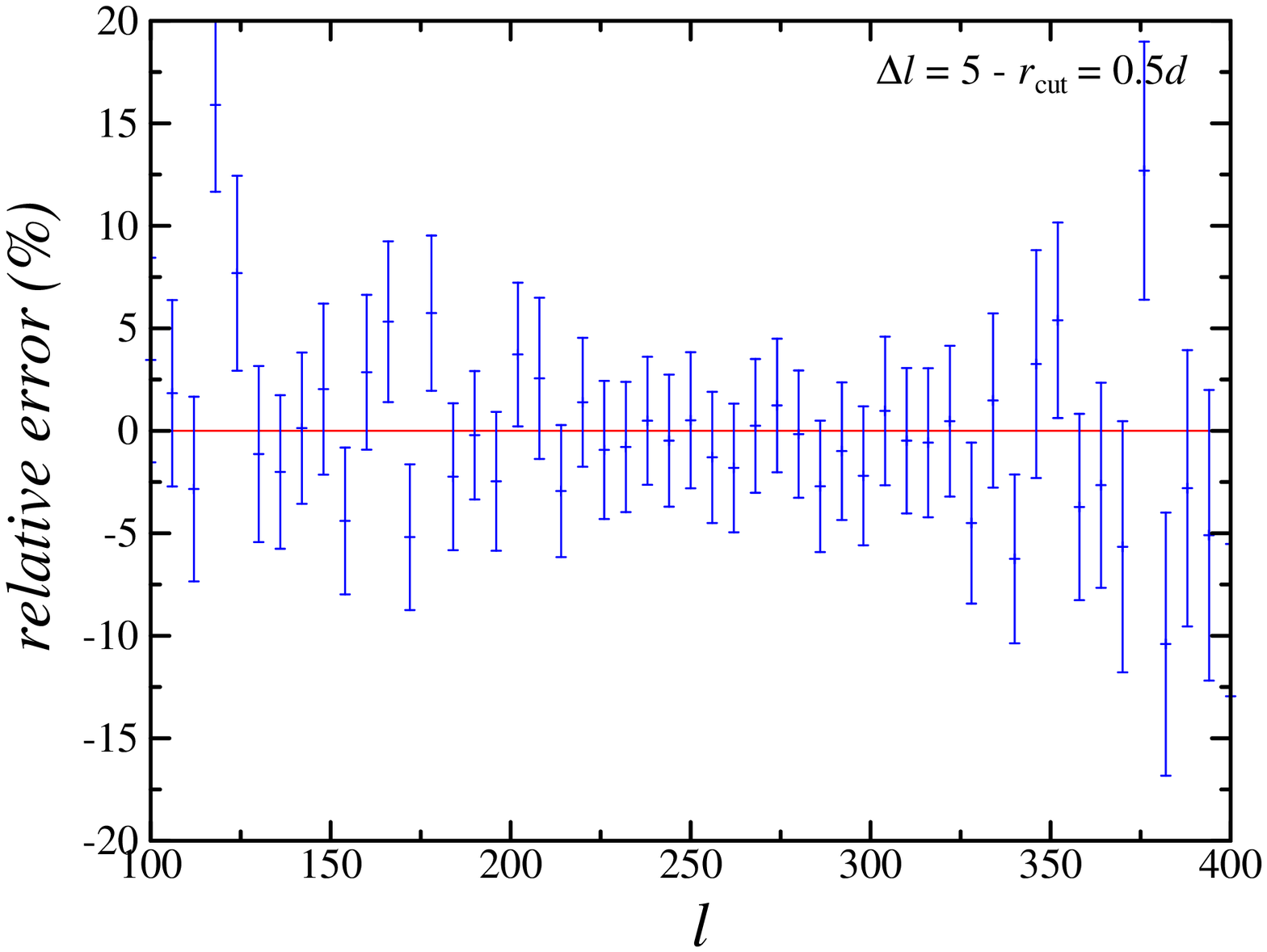}
\includegraphics[width=8cm]{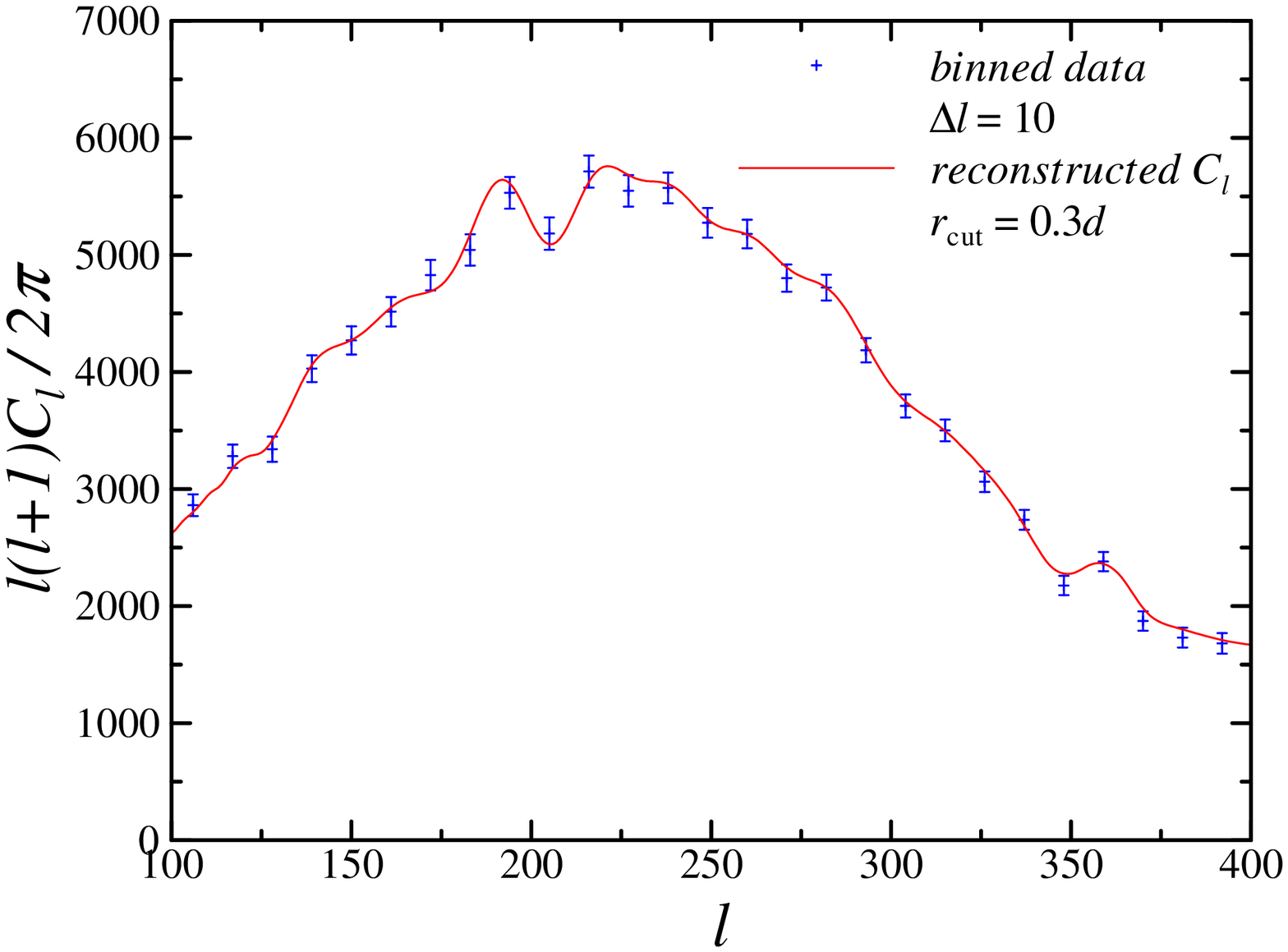}
\includegraphics[width=7cm]{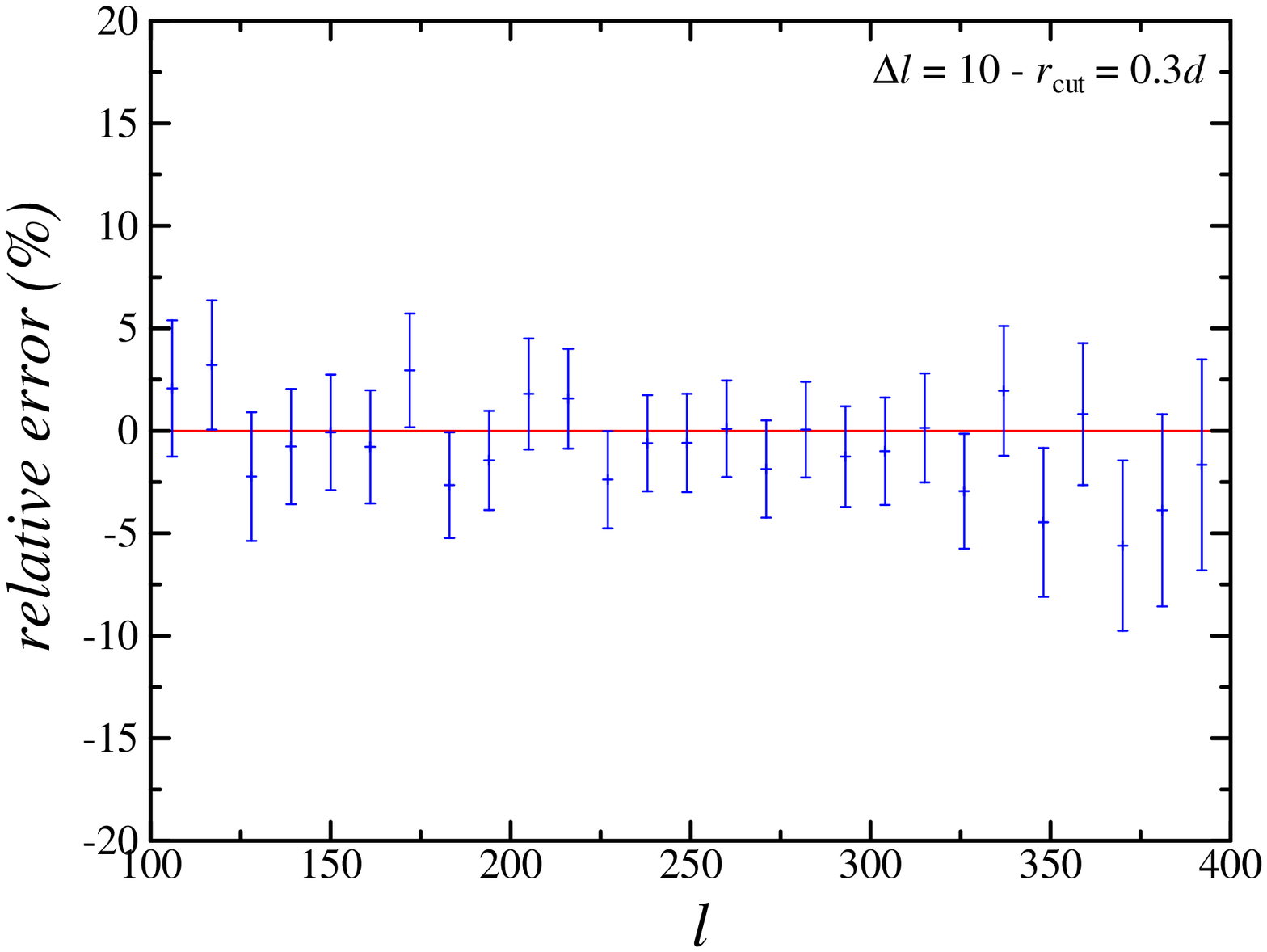}
\caption{Accuracy check of our reconstruction method. 
We show the cases of 
$r_{{\rm cut}} \simeq 0.5d$, which leads to 
$\Delta\ell \sim 6$ ({\it top}), 
and $r_{{\rm cut}} \simeq 0.3d$, which leads to 
$\Delta\ell \sim 10$ ({\it bottom}), 
from the relation $\Delta\ell \sim \Delta kd \simeq \pi d/r_{{\rm cut}}$. 
{\it Left}: Comparison of the binned WMAP data, 
$C^{{\rm bin}}_\ell$ ({\it plus signs with error bars}), 
with the angular power spectrum, 
$C^{{\rm re}}_\ell$ ({\it solid curve}), 
recovered from the reconstructed $P(k)$ shown in Fig.~\ref{CUT}. 
{\it Right}: Relative errors, 
$(C^{{\rm bin}}_\ell-C^{{\rm re}}_\ell)/C^{{\rm re}}_\ell$. 
The relative errors are small for most of the bins 
except for those corresponding to the scales of 
the singularities. \label{RECL}}
\end{center}
\end{figure}

We show $P(k)$ reconstructed from the WMAP data 
for the fiducial cosmological parameter set, namely, 
$h=0.72$, $\Omega_b=0.047$, $\Omega_\Lambda=0.71$, 
$\Omega_m=0.29$, and $\tau=0.17$ in Fig.~\ref{REWMAP}. 
We can see oscillations 
whose amplitude is about $20\%-30\%$ of the mean value 
with frequency $(\Delta kd)^{-1} \simeq 1/15-1/10$. 
To check whether our method works correctly, 
we recalculate $C_\ell$ from the obtained $P(k)$ 
in the range $120 \lesssim kd \lesssim 380$, 
assuming scale-invariance outside of this range, 
and compare it with the observational data of WMAP. 
As mentioned in Sec.~\ref{METHOD}, 
our method reconstructs $P(k)$ with a finite resolution, 
which is caused by the cutoff scale of the Fourier sine integral. 
In Fig.~\ref{CUT}, we compare the cases of 
$r_{{\rm cut}} \simeq 0.5d$ and $\simeq 0.3d$, 
which lead to $\Delta \ell \sim \Delta kd \simeq 6$ or 
$\Delta k \simeq 4.5\times10^{-4}{\rm Mpc}^{-1}$, 
and $\Delta \ell \sim \Delta kd \simeq 10$ or 
$\Delta k \simeq 7.5\times10^{-4}{\rm Mpc}^{-1}$, 
respectively. 
We find that the recalculated $C_\ell$ spectra agree with 
the binned WMAP data corresponding to the respective smoothing scales 
as shown in Fig.~\ref{RECL}. 
These agreements are quite impressive. 
We also note that the characteristic frequency of oscillation 
changes accordingly as we vary the cutoff scale, 
so the observed oscillatory behavior with 
frequency $(\Delta kd)^{-1} \simeq 1/15-1/10$ 
in Fig.~\ref{REWMAP} does not necessarily have fundamental meaning. 
We confirm that the resolution becomes better 
as the cutoff scale is made larger, 
in agreement with the relation $\Delta k \simeq \pi/r_{{\rm cut}}$. 
Of course, we adopt the former resolution, 
the finest possible scheme without numerical instability 
in subsequent sections. 

\subsection{Reconstruction from Binned Data} \label{BINNED}
\begin{figure}
\begin{center}
\includegraphics[width=9cm]{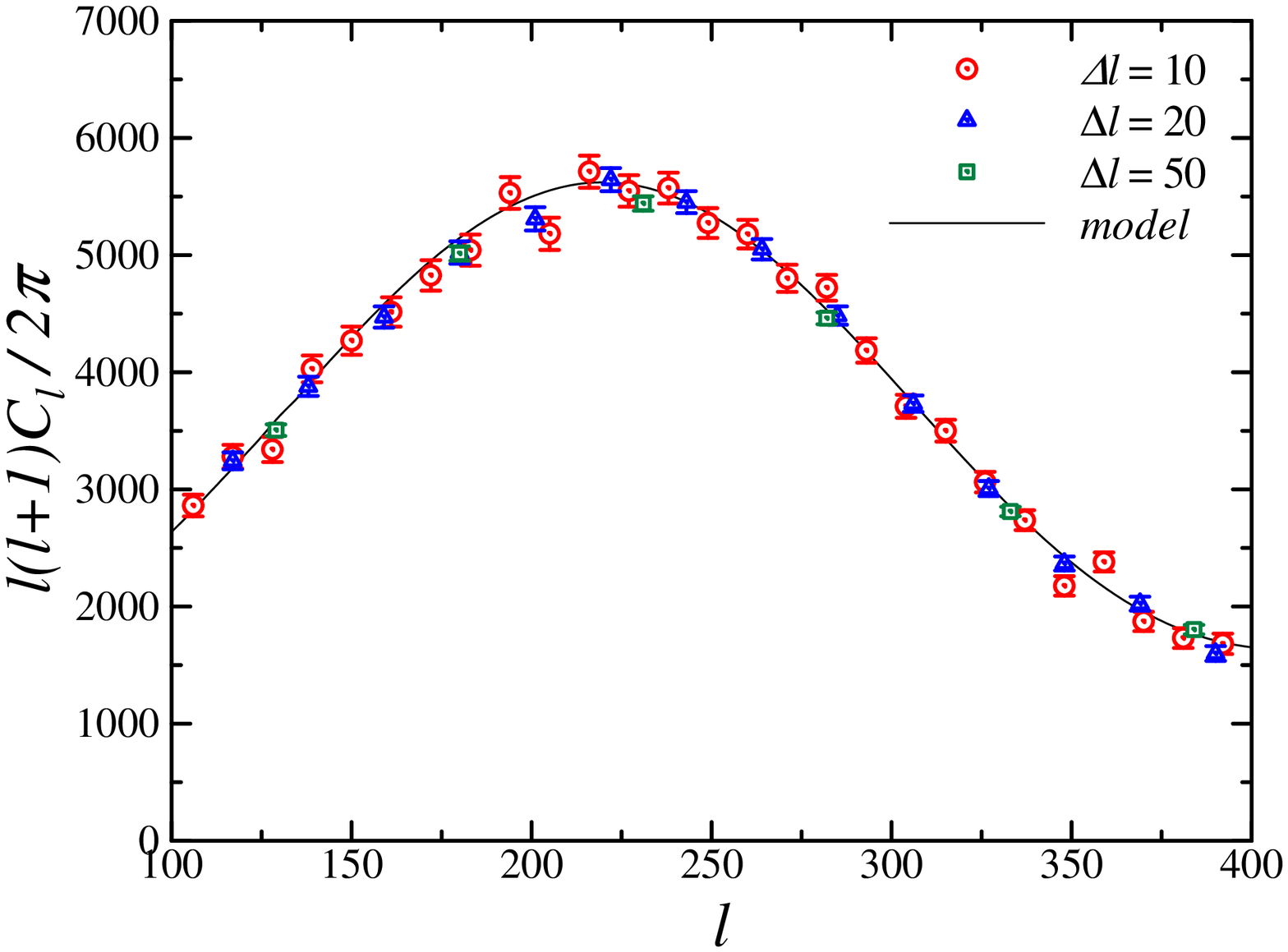}
\includegraphics[width=9cm]{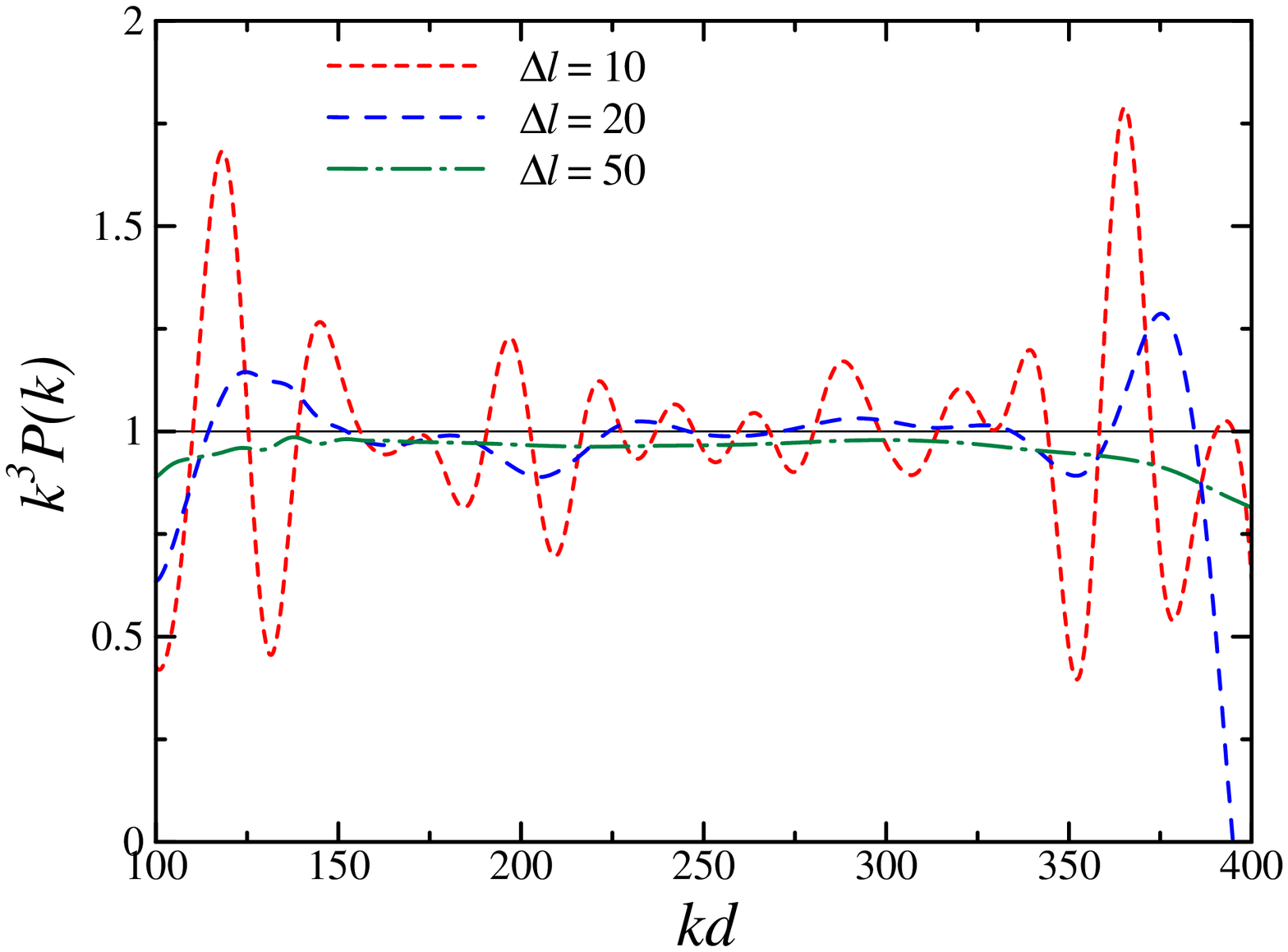}
\caption{Reconstruction from the binned WMAP data. 
{\it Top}: Binned WMAP data whose bin sizes are $\Delta\ell=10, 20,$ and $50$; 
the solid curve represents 
the fiducial model for the scale-invariant spectrum. 
{\it Bottom}: The $P(k)$ spectra reconstructed from the binned WMAP data. 
As the bin size is made smaller, we can see more oscillations. \label{BINWMAP}}
\end{center}
\end{figure}

Figure~\ref{BINWMAP} demonstrates the dependence of the reconstructed $P(k)$ 
on the bin size of the data. 
These $P(k)$ spectra are reconstructed from the binned WMAP data 
whose bin sizes are $\Delta\ell=10, 20,$ and $50$, 
by interpolating through the binned data points, respectively. 
This is to see both global and local features in $P(k)$. 
It is found that the oscillatory feature becomes more prominent 
as the bin size is made smaller, 
even if the spectrum globally looks scale-invariant. 
We emphasize that such nontrivial features cannot be quantified in $P(k)$ 
as long as conventional parameter-fitting methods are used. 

\subsection{Statistical Analysis} \label{ANALYSIS}
\begin{figure}
\begin{center}
\includegraphics[width=8cm]{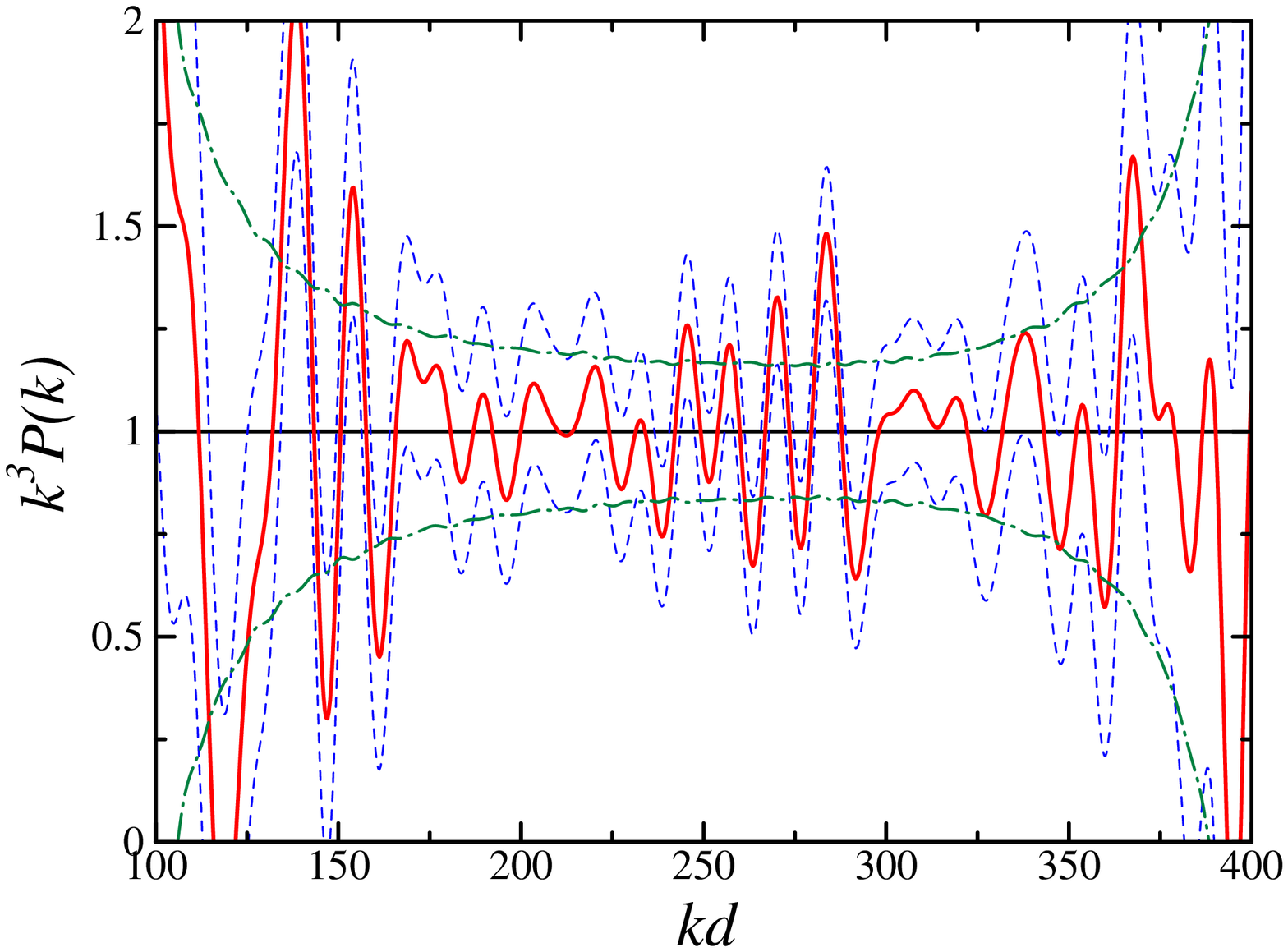}
\includegraphics[width=8cm]{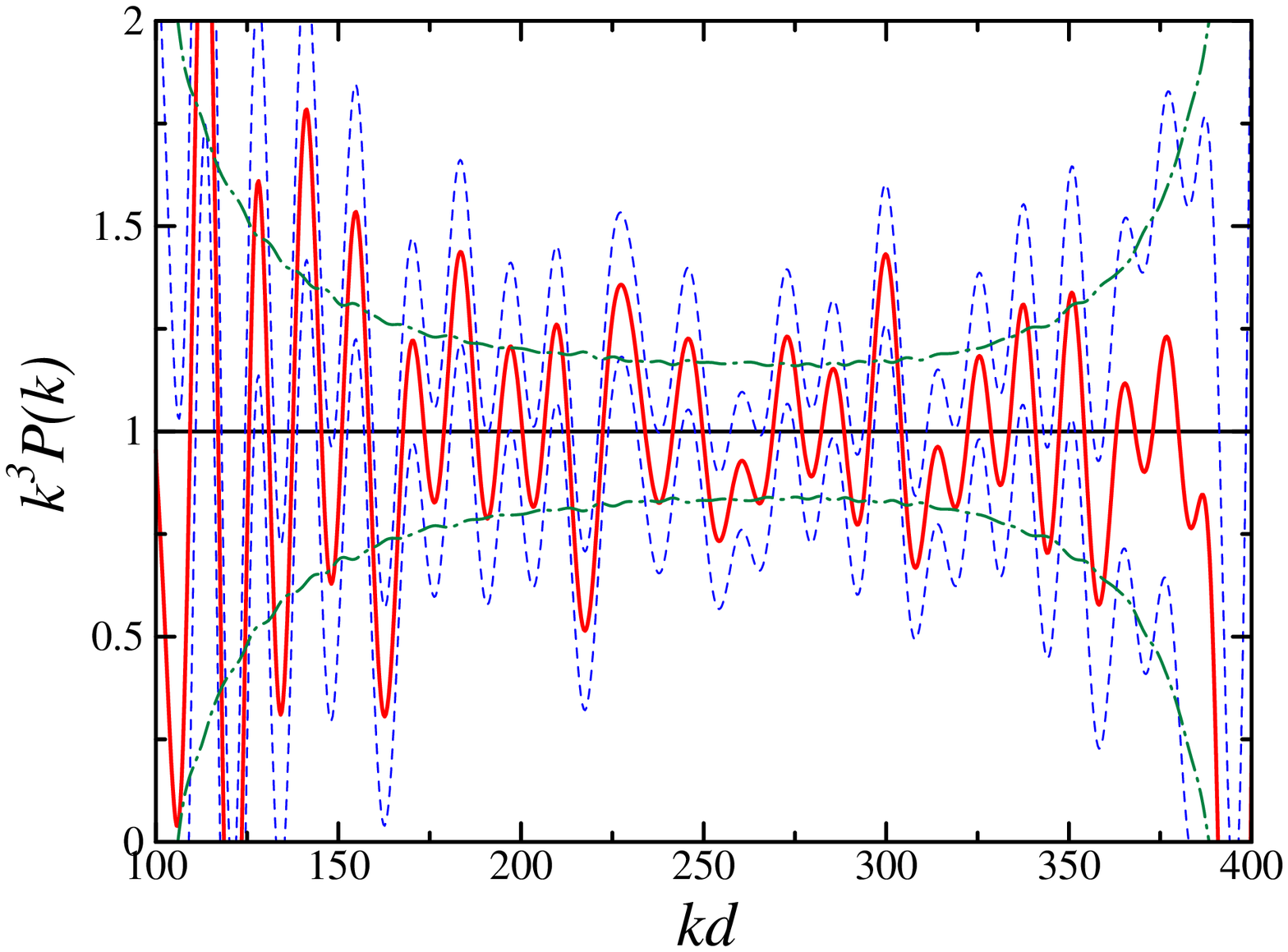}
\includegraphics[width=8cm]{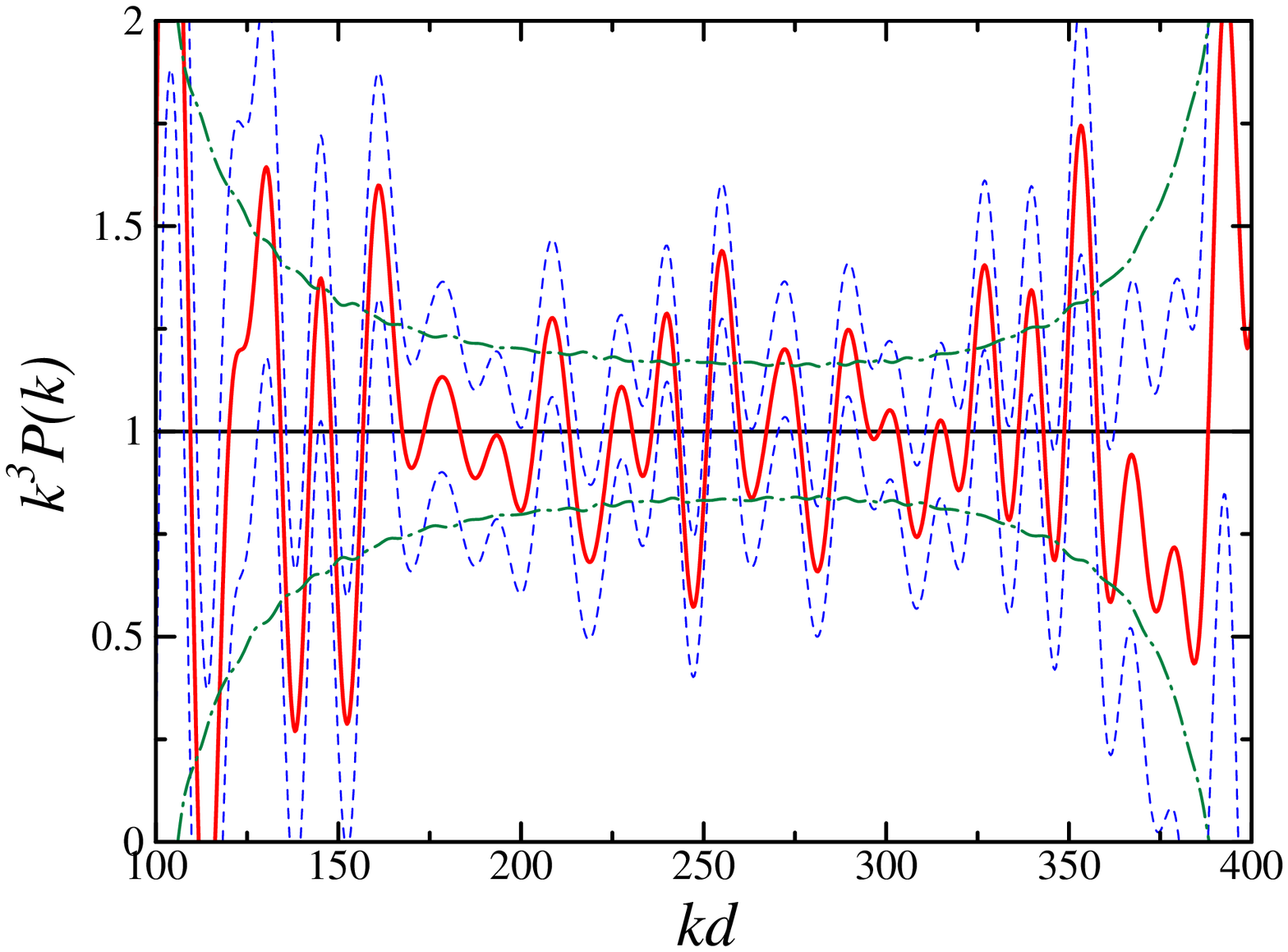}
\includegraphics[width=8cm]{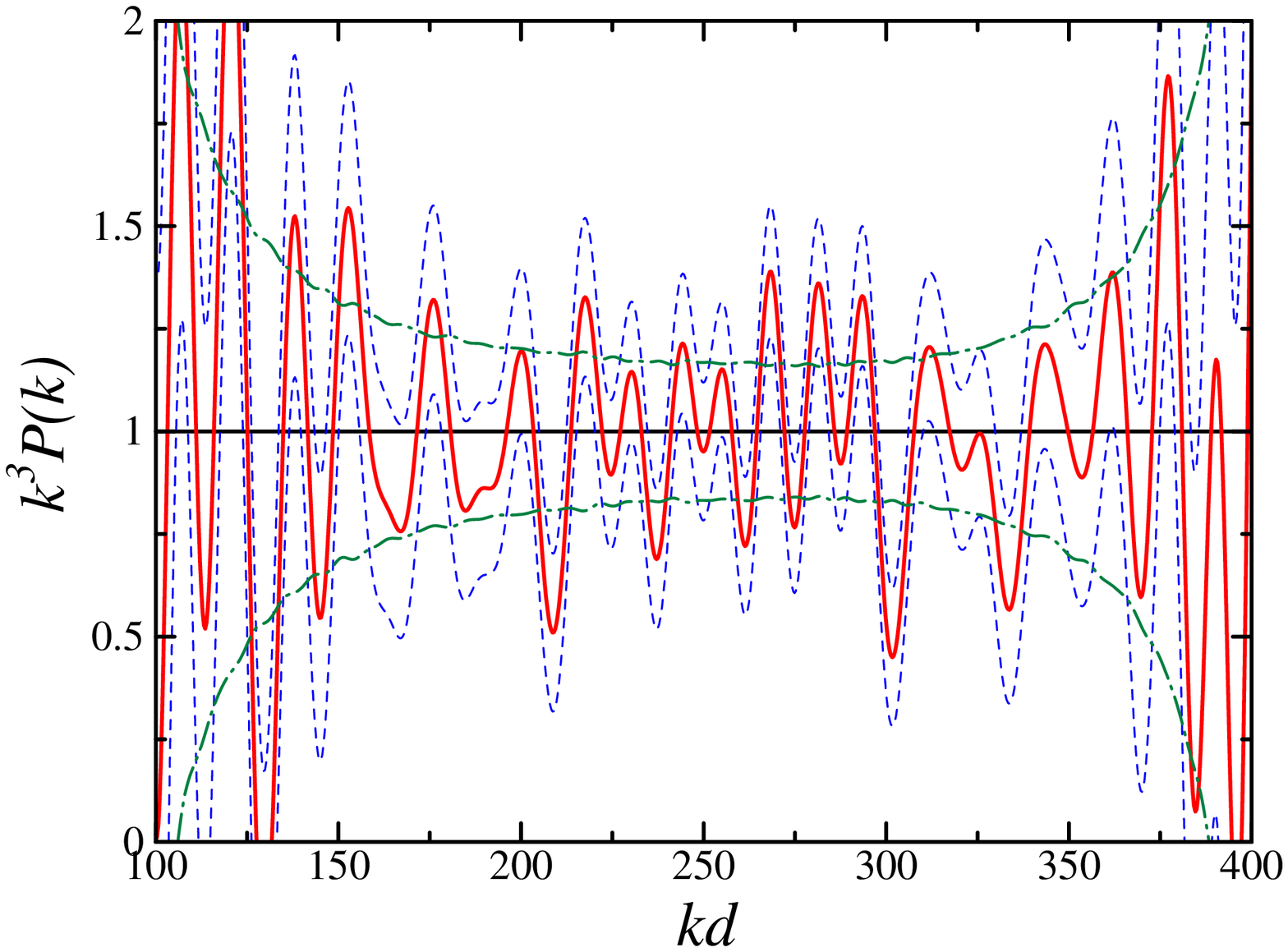}
\caption{Primordial spectra $P(k)$ reconstructed 
from artificial CMB data for four different realizations. 
The original spectrum is taken to be scale invariant, 
and each realization is generated 
by drawing a random number to each $C_\ell$ 
with the same error as the WMAP data, 
assuming that the cosmological parameters are known. 
The same oscillatory features as shown in Fig.~\ref{REWMAP} 
are seen. \label{RESIM}}
\end{center}
\end{figure}

To examine whether the oscillatory features are real, 
we perform the following simulations and compare the results. 
First, we assume a scale-invariant $P(k)$ 
and calculate theoretical $C_\ell$ spectra 
for the fiducial cosmological parameter set. 
Then, we make artificial data of $C_\ell$ spectra 
with the same errors as those of the WMAP data 
by drawing random numbers. 
We use these artificial data to reconstruct $P(k)$, 
assuming that the cosmological parameters are known. 
The resultant $P(k)$ spectra for some different realizations 
are shown in Fig.~\ref{RESIM}. 
We can see that these reconstructed spectra also have oscillatory features 
whose amplitude and frequency are almost the same as 
the $P(k)$ from the WMAP data. 
This is caused purely by the scatter of the data. 
In other words, even if $P(k)$ is really scale-invariant, 
it is likely that the reconstructed $P(k)$ looks oscillatory 
because of the observational errors, 
and thus it is difficult to conclude whether or not 
there are some significant features 
in the reconstructed $P(k)$ from the WMAP data. 
It is necessary to quantify their significance. 

For this purpose, 
we define the deviation from the scale-invariance in $P(k)$ 
in the range between $k_1$ and $k_2$ as 
\begin{eqnarray}
D(k_1,k_2) \equiv \int_{k_1}^{k_2} \! dk \, \left[ k^3P(k)-A \right]^2,
\label{DEVI}
\end{eqnarray}
and evaluate their statistical significance as follows. 
First, we calculate $D(k_1,k_2)$ for each reconstructed $P(k)$ 
from mock $C_\ell$ data for the scale-invariant $P(k)$ 
in the same simulation as mentioned above, 
with the fiducial cosmological parameters. 
Then, we estimate the probability that 
$D(k_1,k_2)$ for the simulation exceeds its observed value 
in the same range of $k$, 
by counting such events for 10,000 realizations. 
We show the results of this analysis in Table~\ref{SIG}. 
It is found that only 1.41\% of the simulations exceed the observed value 
in the range $100 \le kd \le 400$, 
and especially, only 0.39\% in the range $175 \le kd \le 225$ 
and 0.59\% in the range $325 \le kd \le 375$, 
from which we may conclude that 
there are some possible deviations around $kd \simeq 200$ and $\simeq 350$, 
as we can also see some prominent features in Fig.~\ref{REWMAP}. 
This is consistent with \citet{WMAPPARA}, 
who show the contribution to $\chi^2$ per multipole in $C_\ell$ 
after fitting the $\Lambda$CDM model with the power law $P(k)$ to the data. 

\subsection{Cosmological Parameters} \label{PARAMETERS}
As is seen in Fig.~\ref{REWMAP}, 
the reconstructed $P(k)$ exhibits severe oscillations, 
reaching negative values locally around the singularity $kd \simeq 430$. 
One may wonder whether this is due to an inappropriate choice of 
the fiducial cosmological parameters, 
because if the different cosmological parameters from actual values 
are adopted in the inversion, 
the reconstructed $P(k)$ is distorted from the real one, 
especially around the singularities as shown by \citet{MSY03}. 
We find, however, 
that 39.36\% of the simulations based on the scale-invariant spectrum 
with the cosmological parameters fixed to the fiducial values 
result in larger values of $D(k_1,k_2)$ in the range $380 \le kd \le 430$. 
Hence, it is likely that the severe oscillations observed in Fig.~\ref{REWMAP} 
are not caused by an inappropriate choice of the cosmological parameters 
but are simply due to large observational errors. 
Still, it is important to see the dependence of the reconstructed $P(k)$ 
on the cosmological parameters, for the oscillatory features may change 
and we may be able to find a better choice of their values. 
In the ideal situations with sufficiently small observational errors, 
we may even constrain the cosmological parameters 
by requiring that the reconstructed $P(k)$ be positive definite. 

\begin{figure}
\begin{center}
\includegraphics[width=8cm]{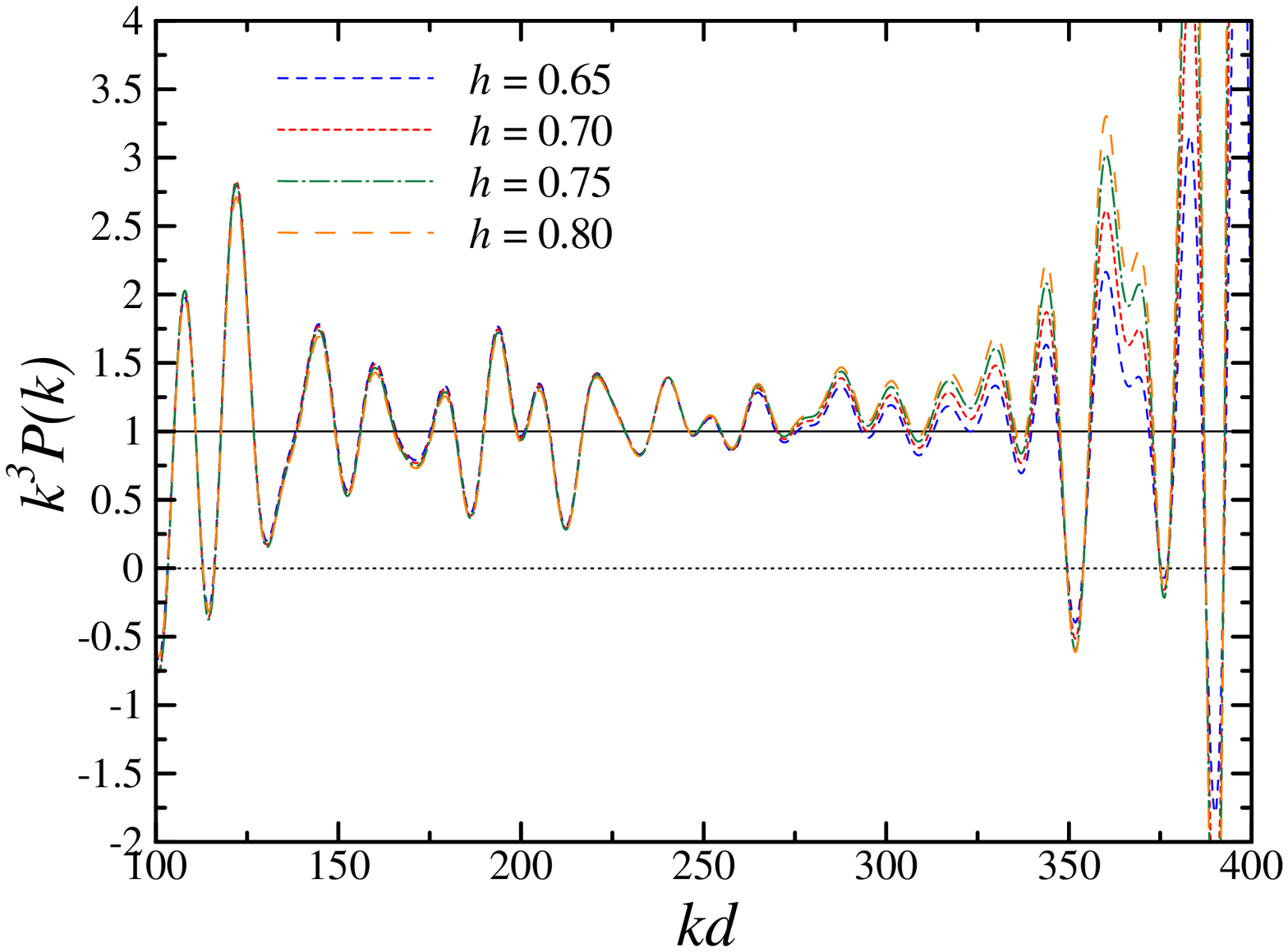}
\includegraphics[width=8cm]{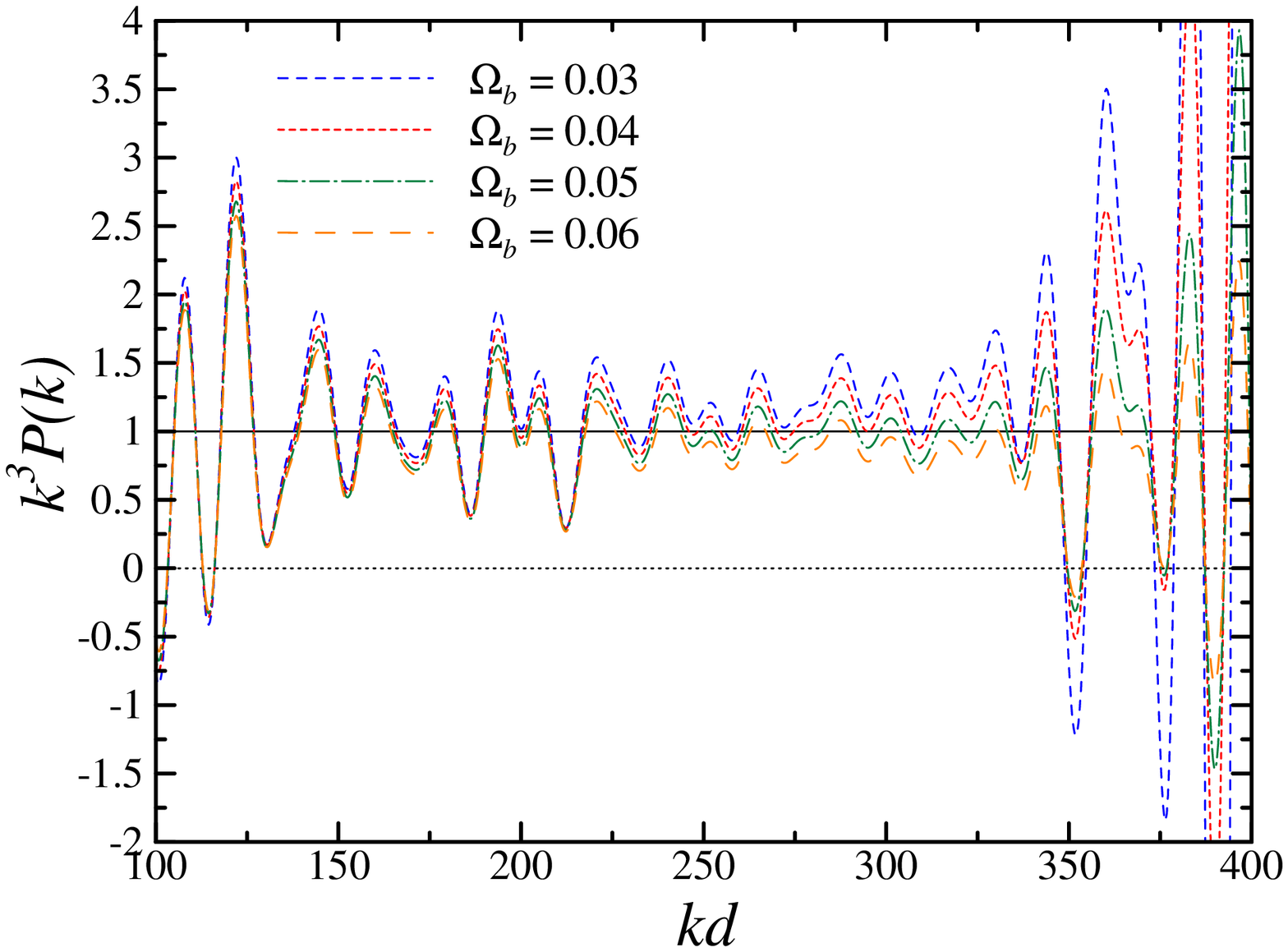}
\includegraphics[width=8cm]{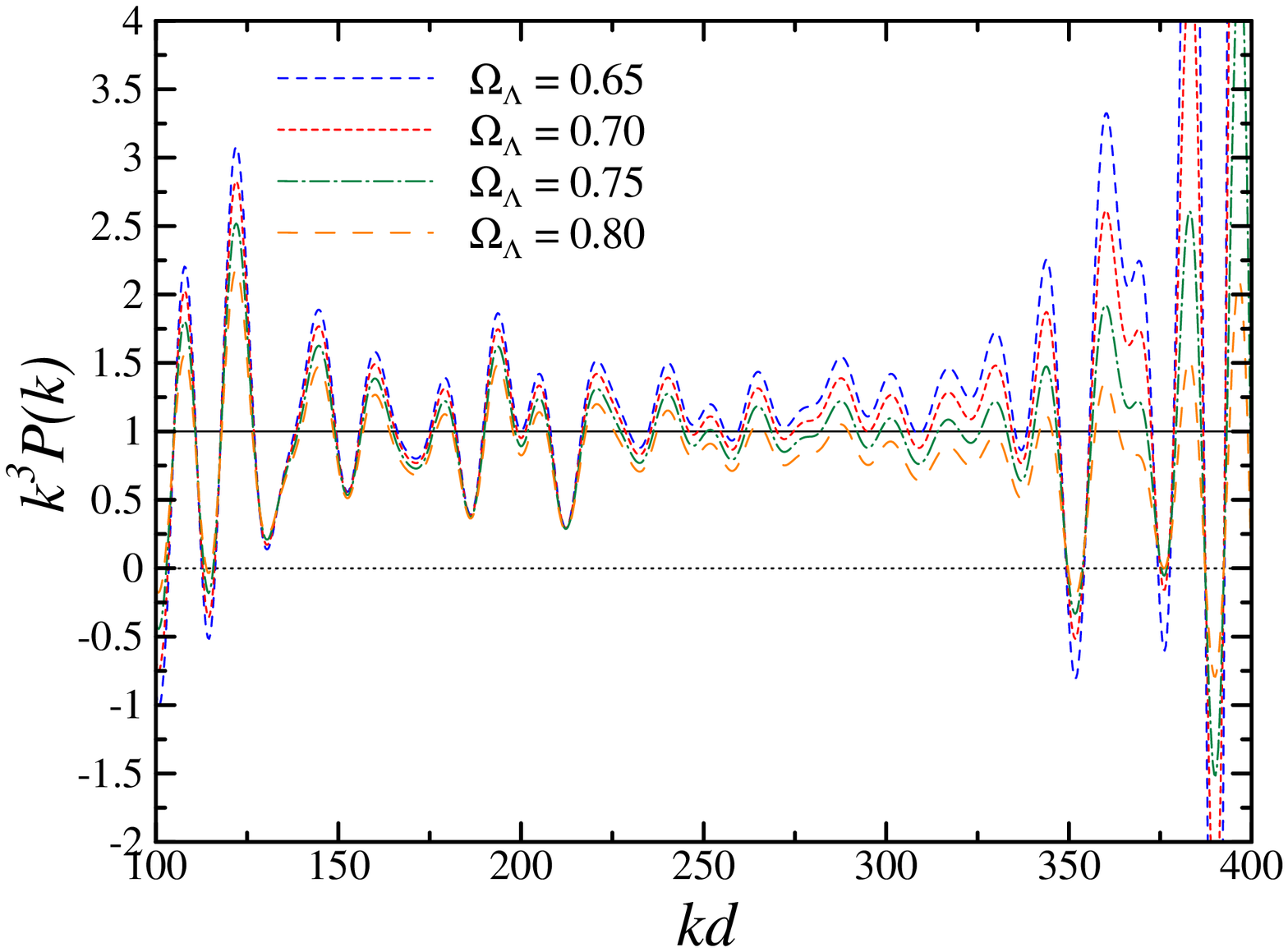}
\caption{Some cases of different cosmological parameter sets. 
We show the reconstructed $P(k)$ from the WMAP data for 
$h=0.65, 0.70, 0.75, 0.80$, $\Omega_b=0.04$, $\Omega_\Lambda=0.70$ 
({\it top left panel}), 
$h=0.70$, $\Omega_b=0.03, 0.04, 0.05, 0.06$, $\Omega_\Lambda=0.70$ 
({\it top right panel}), 
and $h=0.70$, $\Omega_b=0.04$, $\Omega_\Lambda=0.65, 0.70, 0.75, 0.80$ 
({\it bottom panel}), 
respectively. 
The oscillatory features are roughly similar to each other 
throughout this range, 
but they are amplified around the second singularity 
in some cases. \label{PARA}}
\end{center}
\end{figure}

Figure~\ref{PARA} depicts the dependence of the reconstructed $P(k)$ 
on the cosmological parameters. 
It is found that the oscillatory features remain, 
but the global amplitude and especially the sharpness of the oscillations 
around the second singularity change. 
We also find some degree of degeneracy between 
$h$, $\Omega_b$, and $\Omega_\Lambda$. 
In Fig.~\ref{PARA}, 
we can see that the increase in $h$ has a similar effect 
as the decrease in $\Omega_b$ or $\Omega_\Lambda$, 
that is, the amplitude of the oscillations increases similarly. 
This degeneracy, however, may be resolved if we could observe fine structures. 
For example, the reconstructed $P(k)$ is practically insensitive to $h$ 
in the range $kd \lesssim 250$; 
hence, variation in $h$ may be distinguished from 
that of $\Omega_b$ or $\Omega_\Lambda$. 

With the current magnitude of observational errors in $C_\ell$, 
we encounter negative values of the reconstructed $P(k)$ 
for the scale-invariant spectrum. 
Hence, we need $C_\ell$ with smaller errors 
to apply this positivity criterion directly. 
In any case, we should examine $P(k)$ systematically 
for as many parameter sets as possible 
beyond those shown in Fig.~\ref{PARA}, 
and we need a statistical method to quantify the positivity criterion 
in the presence of observational errors. 
This issue is currently under study. 

At present, we can reconstruct $P(k)$ only in the range 
$100 \lesssim kd \lesssim 400$.
If $C_\ell$ data improve at $\ell \gtrsim 700$, 
we will be able to obtain $P(k)$ up to the third singularity, $kd \simeq 680$. 
This additional information on $P(k)$ around the third singularity will be 
important for constraining the cosmological parameters, 
partly because of reduction in observational errors as discussed above 
and partly because the shape of $P(k)$ around the singularities 
is sensitive to the cosmological parameters, 
as mentioned in Sec.~\ref{ANALYSIS}. 

\section{Conclusion} \label{CONCLUSION}
We have reconstructed the shape of the primordial spectrum, $P(k)$, 
from the WMAP angular power spectrum data, $C_\ell$, 
by the inversion method proposed by \citet{MSY02,MSY03}, 
under the assumptions that the universe is spatially flat 
and the primordial fluctuations are purely adiabatic. 
First, we have set the cosmological parameters as 
$h=0.72$, $\Omega_b=0.047$, $\Omega_\Lambda=0.71$, 
$\Omega_m=0.29$, and $\tau=0.17$. 
We have obtained an oscillatory $P(k)$. 
The amplitude of oscillations is about $20\%-30\%$ of the mean value, 
and the frequency is about $(\Delta kd)^{-1} \simeq 1/15-1/10$, 
which reflects the resolution of our method. 
We have also confirmed that the reconstructed $P(k)$ 
has good accuracy in the range $120 \lesssim kd \lesssim 380$, 
or $9.0\times10^{-3}{\rm Mpc}^{-1} \lesssim k 
\lesssim 2.8\times10^{-2}{\rm Mpc}^{-1}$, 
with a resolution of 
$\Delta kd \simeq 5$, or $\Delta k \simeq 3.7\times10^{-4}{\rm Mpc}^{-1}$. 
Thus, our inversion method can reconstruct $P(k)$ 
with a much finer resolution than other methods proposed so far, 
such as the binning, wavelet band powers, or direct wavelet expansion method 
\citep{WSS99a,WSS99b,SH01,SH03,WM02,BLWE03,MW03A,MW03B,MW03C}. 

To examine the statistical significance of possible nontrivial features 
in the reconstructed spectrum, 
we have generated 10,000 sets of mock CMB data 
from the scale-invariant spectrum 
and reconstructed $P(k)$ from these artificial data. 
We have found that only 1.41\% of them have the values of $D(k_1,k_2)$ 
larger than the observed value in the range $100 \le kd \le 400$, 
or $7.5\times10^{-3}{\rm Mpc}^{-1} \lesssim k 
\lesssim 3.0\times10^{-2}{\rm Mpc}^{-1}$, 
only 0.39\% in the range $175 \le kd \le 225$, 
and 0.59\% in the range $325 \le kd \le 375$. 
From these results we conclude that there are some possible deviations 
from scale-invariance around 
$kd \simeq 200$, or $k \simeq 1.5\times10^{-2}{\rm Mpc}^{-1}$, 
and $kd \simeq 350$, or $k \simeq 2.6\times10^{-2}{\rm Mpc}^{-1}$. 
On the other hand, we conclude that 
the severe oscillation of the reconstructed $P(k)$ around the singularity 
$kd \simeq 430$, which drives $P(k)$ to negative values in some regions, 
is due not to an inappropriate choice of the cosmological parameters 
but to the large observational errors. 
This is because we find a high probability around the singularity 
$kd \simeq 430$, or 39.36\% in the range $380 \le kd \le 430$. 

There are some issues that we plan to investigate in detail. 
First, it will be interesting to perform the reconstruction 
for a much wider range of the cosmological parameters 
and obtain systematic constraints on them, 
as mentioned in Sec.~\ref{PARAMETERS}. 
Second, it may be necessary to 
apply the $D(k_1,k_2)$ test to other smooth forms of 
$P(k)$, for example, to those reconstructed from the binned data 
instead of the scale-invariant $P(k)$, as discussed in Sec.~\ref{BINNED}, 
for various values of the cosmological parameters. 
This is because the global shape of the reconstructed $P(k)$ changes 
depending on the cosmological parameters. 
Third, we should extend our formalism to 
include the CMB polarization so that it may be applied not only to 
the WMAP data but also to future CMB observations, such as 
Planck.\footnote{See http://www.rssd.esa.int/index.php?project=PLANCK} 
Inclusion of the CMB polarization will constrain $P(k)$ more severely, 
and if the B-mode polarization is detected, 
we can investigate the tensor mode of the primordial perturbations 
\citep{AAS79,RSV82,AW84,AGP85,CBDES93,CDS93}. 

\acknowledgements
We would like to thank Eiichiro Komatsu for fruitful discussions. 
This work was supported in part by JSPS Grants-in-Aid for 
Scientific Research 12640269 (M.S.) and 13640285 (J.Y.) 
and by Monbu-Kagakusho Grant-in-Aid for 
Scientific Research (S) 14102004 (M.S.). 
N.K. is supported by Research Fellowships of JSPS 
for Young Scientists (04249).

\clearpage

\clearpage

\clearpage

\clearpage

\clearpage

\clearpage

\begin{table}
\begin{center}
\caption{Statistical Significance of the Deviations from the Scale Invariance 
\label{SIG}}
\vspace{1cm}
\begin{tabular}{c|cccccc}
\tableline\tableline
range $[k_1d,k_2d]$ & [100,150] & [150,200] & [200,250] 
                    & [250,300] & [300,350] & [350,400] \\
\tableline
probability & 18.20\% &  9.72\% &  3.08\% 
            & 83.31\% & 16.52\% &  2.23\% \\
\tableline
\end{tabular}

\vspace{1cm}
\begin{tabular}{c|cccc}
\tableline\tableline
range $[k_1d,k_2d]$ & [100,400] & [175,225] & [325,375] & [380,430] \\
\tableline
probability &  1.41\% &  0.39\% &  0.59\% & 39.36\% \\
\tableline
\end{tabular}
\tablecomments{We show the probabilities that 
the deviation $D(k_1,k_2)$ for the simulation exceeds its observed value 
in various ranges $[k_1d,k_2d]$. 
The simulations are performed for the scale-invariant spectrum 
with the fiducial cosmological parameters, which are assumed to be known. }
\end{center}
\end{table}

\clearpage

\begin{thebibliography}{}

\bibitem[Abbott and Wise(1984)]{AW84}
Abbott, L. F., and Wise, M. 1984, Nucl. Phys., B244, 541

\bibitem[Bennett et al.(1996)]{COBE}
Bennett, C. L., Banday, A. J., Gorski, K. M., Hinshaw, G., Jackson, P., 
Keegstra, P., Kogut, A., Smoot, G. F., Wilkinson, D. T., Wright, E. L. 
1996, \apjl, 464, L1

\bibitem[Bennett et al.(2003)]{WMAPBASIC}
Bennett, C. L., Halpern, M., Hinshaw, G., Jarosik, N., Kogut, A., Limon, M., 
Meyer, S. S., Page, L., Spergel, D. N., Tucker, G. S., Wollack, E., 
Wright, E. L., Barnes, C., Greason, M. R., Hill, R. S., Komatsu, E., 
Nolta, M. R., Odegard, N., Peirs, H. V., Verde, L., Weiland J. L. 
2003, \apjs, 148, 1

\bibitem[Bi et al.(2003)]{BFZ03}
Bi, X., Feng, B., and Zhang, X., hep-ph/0309195

\bibitem[Bridle et al.(2003)]{BLWE03}
Bridle, S. L., Lewis, A. M., Weller, J., and Efstathiou, G. 
2003, Mon. Not. Roy. Astron. Soc., 342, L72

\bibitem[Chung et al.(2003)]{CST03}
Chung, D. J. H., Shiu, G., and Trodden, M. 2003, \prd, 68, 063501

\bibitem[Cline et al.(2003)]{CCL03}
Cline, J. M., Crotty, P., and Lesgourgues, J. 2003, JCAP, 0309, 010

\bibitem[Contaldi et al.(2003)]{CPKL03}
Contaldi, C. R., Peloso, M., Kofman, L., and Linde, A. 2003, JCAP, 0307, 002

\bibitem[Crittenden et al.(1993a)]{CBDES93}
Crittenden, R., Bond, J. R., Davis, R. L., Efstathiou, G., 
and Steinhardt, P. J. 1993a, \prl, 71, 324

\bibitem[Crittenden et al.(1993b)]{CDS93}
Crittenden, R., Davis, R. L., and Steinhardt, P. J. 1993b, \apjl, 417, L13

\bibitem[de Deo et al.(2003)]{DCS03}
de Deo, S., Caldwell, R. R., and Steinhardt. P. J. 2003, \prd, 67, 103509

\bibitem[de Oliveira-Costa et al.(2003)]{DTZH03}
de Oliveira-Costa, A., Tegmark, M., Zaldarriaga, M., and Hamilton, A., 
astro-ph/0307282

\bibitem[Dvali and Kachru(2003)]{DK03}
Dvali, G., and Kachru, S., hep-ph/0310244

\bibitem[Efstathiou(2003a)]{GE03A}
Efstathiou, G. 2003a, Mon. Not. Roy. Astron. Soc., 343, L95

\bibitem[Efstathiou(2003b)]{GE03B}
Efstathiou, G. 2003b, Mon. Not. Roy. Astron. Soc., 346, L26

\bibitem[Efstathiou(2004)]{GE04}
Efstathiou, G. 2004, Mon. Not. Roy. Astron. Soc., 348, 885

\bibitem[Feng et al.(2003)]{FLZZ03}
Feng, B., Li, M., Zhang, R. J., and Zhang, X. 2003, \prd, 68, 103511

\bibitem[Feng and Zhang(2003)]{FZ03}
Feng, B., and Zhang, X. 2003, Phys. Lett., B570, 145

\bibitem[Gazta${\rm \tilde{n}}$aga et al.(2003)]{GWMMH03}
Gazta${\rm \tilde{n}}$aga, E., Wagg, J., Multam${\rm \ddot{a}}$ki, T., 
Monta${\rm \tilde{n}}$a, A., and Hughes, D. H. 
2003, Mon. Not. Roy. Astron. Soc., 346, L47

\bibitem[Hannestad(2001)]{SH01}
Hannestad, S. 2001, \prd, 63, 043009

\bibitem[Hannestad(2003)]{SH03}
Hannestad, S., astro-ph/0311491

\bibitem[Hu and Sugiyama(1995)]{HS95}
Hu, W., and Sugiyama, N.  1995, \apj, 444, 489

\bibitem[Huang and Li(2003)]{HL03}
Huang, Q. G., and Li, M. 2003, High Energy Phys., 06, 014

\bibitem[Kawasaki et al.(2003)]{KYY03}
Kawasaki, M., Yamaguchi, M., and Yokoyama, J. 2003, \prd, 68, 023508

\bibitem[Kawasaki and Takahashi(2003)]{KT03}
Kawasaki, M., and Takahashi, F. 2003, Phys. Lett., B570, 151

\bibitem[Kodama and Sasaki(1984)]{KS84}
Kodama, H., and Sasaki, M.  1984, Prog. Theor. Phys. Suppl., 78, 1

\bibitem[Komatsu et al.(2003)]{WMAPGAUSS}
Komatsu, E., Kogut, A., Nolta, M., Bennett, C. L., Halpern, M., Hinshaw, G., 
Jarosik, N., Limon, M., Meyer, S. S., Page, L., Spergel, D. N., Tucker, G. S., 
Verde, L., Wollack, E., Wright, E. L. 
2003, \apjs, 148, 119

\bibitem[Kyae and Shafi(2003)]{KS03}
Kyae, B., and Shafi, Q. 2003, JHEP, 0311, 036

\bibitem[Matsumiya et al.(2002)]{MSY02}
Matsumiya, M., Sasaki, M., and Yokoyama, J. 2002, \prd, 65, 083007

\bibitem[Matsumiya et al.(2003)]{MSY03}
Matsumiya, M., Sasaki, M., and Yokoyama, J. 2003, JCAP, 0302, 003

\bibitem[Mukherjee and Wang(2003a)]{MW03A}
Mukherjee, P., and Wang, Y. 2003a, \apj, 593, 38

\bibitem[Mukherjee and Wang(2003b)]{MW03B}
Mukherjee, P., and Wang, Y. 2003b, \apj, 598, 779

\bibitem[Mukherjee and Wang(2003c)]{MW03C}
Mukherjee, P., and Wang, Y. 2003c, \apj, 599, 1

\bibitem[Niarchou et al.(2004)]{NJP04}
Niarchou, A., Jaffe, A. H., and Pogosian, L. 2004, \prd, in press 
(astro-ph/0308461)

\bibitem[Peiris et al.(2003)]{WMAPINF}
Peiris, H. V., Komatsu, E., Verde, L., Spergel, D. N., Bennett, C. L., 
Halpern, M., Hinshaw, G., Jarosik, N., Kogut, A., Limon, M., Meyer, S., 
Page, L., Tucker, G. S., Wollack, E., Wright, E. L.
2003, \apjs, 148, 213

\bibitem[Piao et al.(2003)]{PFZ03}
Piao, Y., Feng, B., and Zhang, X., hep-th/0310206

\bibitem[Polnarev(1985)]{AGP85}
Polnarev, A. G. 1985, Sov. Astron., 29, 607

\bibitem[Rubakov et al.(1982)]{RSV82}
Rubakov, V. A., Sazhin, M. V., and Veryaskin, A. V. 
1982, Phys. Lett., B115, 189

\bibitem[Shafieloo and Souradeep(2003)]{SS03}
Shafieloo, A., and Souradeep, T., astro-ph/0312174

\bibitem[Spergel et al.(2003)]{WMAPPARA}
Spergel, D. N., Verde, L., Peiris, H. V., Komatsu, E., Nolta, M. R., 
Bennett, C. L., Halpern, M., Hinshaw, G., Jarosik, N., Kogut, A., Limon, M., 
Meyer, S. S., Page, L., Tucker, G. S., Weiland, J. L., Wollack, E., 
Wright, E. L. 
2003, \apjs, 148, 175

\bibitem[Starobinsky(1979)]{AAS79}
Starobinsky, A. A. 1979, JETP Lett., 30, 682

\bibitem[Tegmark et al.(2003)]{TDH03}
Tegmark, M., de Oliveira-Costa A., and Hamilton A. 2003, \prd, 68, 123523

\bibitem[Tegmark and Zaldarriaga(2002)]{TZ02}
Tegmark, M., and Zaldarriaga, M. 2002, \prd, 66, 103508

\bibitem[Uzan et al.(2004)]{URLW04}
Uzan, J. P., Riazuelo, A., Lehoucq, R., and Weeks, J. 2004, \prd, 69, 043003

\bibitem[Wang and Mathews(2002)]{WM02}
Wang, Y., and Mathews, G. J. 2002, \apj, 573, 1

\bibitem[Wang et al.(1999a)]{WSS99a}
Wang, Y., Spergel, D. N., and Strauss, M. A. 1999a, 
in AIP Conf. Proc. 478, COSMO-98: 
Second International Workshop on Particle Physics and the Early Universe, 
ed. D. O. Caldwell (Woodbury: AIP), 164

\bibitem[Wang et al.(1999b)]{WSS99b}
Wang, Y., Spergel, D. N., and Strauss, M. A. 1999b, \apj, 510, 20

\bibitem[Yamaguchi and Yokoyama(2003)]{YY03} 
Yamaguchi, M., and Yokoyama, J. 2003, \prd, 68, 123520

\bibitem[Yokoyama(1999)]{JY99}
Yokoyama, J. 1999, \prd, 59, 107303

\end{thebibliography}
\end{document}